\documentclass[10pt,journal,twoside,final]{IEEEtran}

\usepackage{multirow}
\usepackage{graphicx} 
\usepackage{bigstrut}
\usepackage{multirow}
\usepackage{url}

\ifCLASSOPTIONcompsoc
\usepackage[caption=false,font=normalsize,labelfont=sf,textfont=sf]{subfig}
\else
\usepackage[caption=false,font=footnotesize]{subfig}
\fi

\usepackage{color}

\usepackage{cite,amsfonts}
\usepackage[cmex10]{amsmath}
\usepackage[subnum]{cases}
\usepackage{array}
\newcolumntype{L}[1]{>{\raggedright\let\newline\\\arraybackslash\hspace{0pt}}m{#1}}
\newcolumntype{C}[1]{>{\centering\let\newline\\\arraybackslash\hspace{0pt}}m{#1}}
\newcolumntype{R}[1]{>{\raggedleft\let\newline\\\arraybackslash\hspace{0pt}}m{#1}}
\usepackage{amssymb}
\usepackage{cite}

\begin{document}

\title{FewSense, Towards a Scalable and Cross-Domain Wi-Fi Sensing System Using Few-Shot Learning}

\author{Guolin~Yin,
Junqing~Zhang,~\IEEEmembership{Member,~IEEE,}
Guanxiong~Shen,
and Yingying~Chen,~\IEEEmembership{Fellow,~IEEE}
\thanks{Manuscript received xxx; revised xxx; accepted xxx. Date of publication xxx; date of current version xxx. The work was in part supported by National Key Research and Development Program of China under grant ID 2020YFE0200600. The review of this paper was coordinated by xxx. 
(\textit{Corresponding author: Junqing Zhang})}
\IEEEcompsocitemizethanks{\IEEEcompsocthanksitem G. Yin, J.~Zhang and G. Shen are with the Department of Electrical Engineering and Electronics, University of Liverpool, Liverpool, L69 3GJ, United Kingdom. (email: \{Guolin.Yin, Junqing.Zhang, Guanxiong.Shen\}@liverpool.ac.uk)
\IEEEcompsocthanksitem Y. Chen is with Department of Electrical and Computer Engineering, Rutgers University, Piscataway, NJ 08854, US. (email: yingche@scarletmail.rutgers.edu)
}
\thanks{Color versions of one or more of the figures in this paper are available online at http://ieeexplore.ieee.org.}
\thanks{Digital Object Identifier xxx}
}

\maketitle

\begin{abstract}
Wi-Fi sensing can classify human activities because each activity causes unique changes to the channel state information (CSI). Existing WiFi sensing suffers from limited scalability as the system needs to be retrained whenever new activities are added, which cause overheads of data collection and retraining. Cross-domain sensing may fail because the mapping between activities and CSI variations is destroyed when a different environment or user (domain) is involved. This paper proposed a few-shot learning-based WiFi sensing system, named FewSense, which can recognise novel classes in unseen domains with only few samples. Specifically, a feature extractor was pre-trained offline using the source domain data. When the system was applied in the target domain, few samples were used to fine-tune the feature extractor for domain adaptation. Inference was made by computing the cosine similarity. FewSense can further boost the classification accuracy by collaboratively fusing inference from multiple receivers. We evaluated the performance using three public datasets, i.e., SignFi, Widar, and Wiar. The results show that FewSense with five-shot learning recognised novel classes in unseen domains with an accuracy of 90.3\%, 96.5\% ,82.7\% on SignFi, Widar, and Wiar datasets, respectively. Our collaborative sensing model improved system performance by an average of 30\%.

\end{abstract}


\section{Introduction}\label{section:intro}
\subsection{Overview}
\IEEEPARstart{W}{i-Fi} sensing has recently received extensive research interests~\cite{liu2019wireless,nirmal2021deep}. Wi-Fi is ubiquitous as it has  been equipped in many consumable electronics including laptops, smartphones, tablets, wearable devices such as Fitbits, and smart home appliances, to name but a few. 
Various interesting applications have been inspired, including large-scale movements such as human activity recognition~\cite{wang2014eyes,li2021two,wang2018spatial}, fall detection \cite{wang2016rt}, \cite{zhang2019commercial}, and gait recognition\cite{wang2016gait}, \cite{zhang2019widigr}, as well as small-scale movements such as gesture recognition~\cite{li2020wihf,gao2021towards}, sign language recognition\cite{signfi}, \cite{shang2017robust}, and vital sign detection \cite{zeng2020multisense}, \cite{wang2016human}.
These applications are very useful for our everyday life. For example, gesture recognition can be applied in human-computer interaction and smart home to enable smooth control of devices.

Wi-Fi transmissions experience line-of-sight propagation, reflection, refraction, and scattering, which are affected by the environment and the objects within it~\cite{goldsmith2005wireless}.
When a person performs an activity, e.g., walking or hand moving, the activity will cause unique changes to the radio propagation, which can be measured via the channel state information (CSI). Deep learning can be adopted to learn the unique mapping between the CSI patterns and activity types due to its effective feature extraction and classification capability~\cite{krizhevsky2012imagenet}. A deep learning-based sensing protocol involves two stages. In the training stage, a training dataset should be constructed and a deep learning model is trained offline. Specifically, when an activity is performed, the Wi-Fi signals will get perturbed.
 The receiver will estimate the CSI whose variation is affected by the activity. For each activity, many CSI records are collected. The process is repeated for all the activities and the training dataset is established. A deep learning model can then be trained offline, which only needs to be done once.
During the testing stage, based on the captured CSI, the derived deep learning model will be utilized to infer the activity type. 

\subsection{Limitations of Existing Systems}
Although the above deep learning-based sensing systems can recognise human activities with reasonable accuracy, there are still limitations that restrict practical applications.

\subsubsection{Limited Scalability}
A classic deep learning-based sensing system explores the uniqueness of the mapping between the CSI variations and activities. A common limitation of such systems is that they can only handle the same classes/activities as the ones used during the training process. Whenever a new class needs to be added to or an existing class needs to be removed from the well-trained model, the model should be retrained using data of the existing classes and the new classes. The retraining results in three types of overhead:
\begin{itemize}
	\item The storage of the existing training data, which occupies hard disk space and requires extra maintenance. 
	\item The collection of the new data that is time consuming and labour extensive. 
	\item The computational overhead for retraining is quite high when there is a large amount of data, which cannot be completed in real time.
\end{itemize}
However, adjusting including adding and deleting classes is common in practical applications, e.g., a new sign gesture may need to be enrolled. The lack of scalability in the current approach and design makes it difficult to be deployed in real world scenarios.

\subsubsection{Domain Shift}
The radio signal experiences multipath propagation whose characteristics are determined by the environment and the sensing activity. 
The same activity performed in different environments will cause deviated CSI variations.
In addition, the same activity performed by different users may also have various CSI variations because the users may perform the same activity in a slightly different way or their heights/body shapes are different. 
The parameters uncorrelated to the activity can be denoted as a domain~\cite{kang2021context}, e.g., the environment and/or the user. Source and target domains refer to the domains of training and testing stages, respectively.

When the source and target domains are different, there is a domain shift. The obtained CSI will be varied, and the recognition will be impacted. 
However, it is very common to apply Wi-Fi sensing in different environments and/or for different users. 
Existing solutions can be classified into three categories, namely domain adversarial training-based~\cite{jiang2018towards,kang2021context}, transfer learning-based~\cite{CrossSense}, and domain-independent feature-based~\cite{widar} approaches. 
However, these approaches are subject to the following limitations:
\begin{itemize}
	\item The domain adversarial training-based approach requires the training dataset to cover data from many domains but there are numerous possible domains.
	\item The transfer learning-based approach needs extensive data from the target domain, which sometimes may not be possible.
	\item The domain-independent feature-based method relies on special knowledge to design unique features and may also need extra receivers, e.g., body-coordinate velocity profile (BVP) in~\cite{widar}. In addition, it may suffer from high computational costs, which make it impossible for practical use.
\end{itemize}
Hence, a lightweight cross-domain sensing approach is urgently required to reduce the overhead of collecting data from source or target domains as well as eliminate constrains of extra hardware and computational resources.

\subsection{Few-Shot Learning}
Few-shot learning (FSL) has been widely investigated in computer vision for image classification~\cite{vinyals2016matching}. 
A classic deep learning model aims to learn the unique feature of each class and predict the label based on the feature mapping. In contrast, FSL is a meta-learning technique. Instead of learning unique features, it makes the prediction based on the similarity of the feature sets.
Specifically, a feature extractor will be pre-trained using the base set. Then, a support set will be constructed with a few pairs of data and labels.
Finally, a query set contains data whose label is to be inferred. The prediction is made by comparing the similarity between features of support and query sets.

FSL has recently been applied to Wi-Fi sensing as well~\cite{ding2020rf,xiao2021onefi} to address the scalability and cross-domain issues. 
The support set can be flexibly adjusted because very few samples are required for each class, and no cumbersome training is needed.
The performance of FSL relies on the generalisation capability of the feature extractor, which however requires an extensive base set. Hence, the work in~\cite{xiao2021onefi} designed a virtual gesture generation algorithm to transform existing gestures into virtual gestures. The base set can thus be enriched and the data collection overhead can be mitigated.
However, designing the virtual gesture generation algorithm requires sophisticated knowledge which cannot be extended to other tasks straightforwardly.

\subsection{Contributions}
In the light of the initial success of using FSL in designing the virtual gesture generation algorithm~\cite{xiao2021onefi}, in this work, we further employ FSL to address the scalability issue and domain shift challenge faced by many practical Wi-Fi sensing applications.
We demonstrated that the feature extractor can be generalised to a totally different sensing task. Specifically, we exploited a public dataset as the base set, applied the trained feature extractor to a different dataset and achieved good classification performance, which is a different approach from~\cite{xiao2021onefi}. 
In other word, our work aims to eliminate the need to construct an individual base set for every sensing task. 

We adopt metric-based FSL for designing a scalable and cross-domain Wi-Fi sensing system. A CNN backbone is revised from the classic AlexNet architecture as the feature extractor. We design a collaborative sensing scheme to leverage the spatial diversity gain enabled by multiple receivers. The proposed model is evaluated using three public datasets, i.e, SignFi~\cite{signfi}, Widar 3.0~\cite{widar} and Wiar~\cite{guo2019wiar}. We study in-domain sensing, cross-domain sensing as well as cross-dataset sensing. Our results show that the proposed FSL-based sensing system has a better performance over the state-of-the-art works in the literature. Our contributions are summarised as follows:   
\begin{itemize}
\item We propose an FSL-based Wi-Fi sensing system, FewSense, which is capable of novel  class recognition and cross-domain sensing.
Specifically, the proposed system can achieve $76.3\%$ cross-domain accuracy for $76$ novel sign language gestures using only one labelled sample for each novel class from the target domain, when all the data is from the SignFi dataset.
\item FewSense can be fine-tuned to new sensing tasks using a few samples. When the feature extractor is trained on the SignFi dataset (sign language recognition), evaluation over the Widar (gesture recognition) and Wiar (human activity recognition) datasets can achieve an average accuracy of 96.5\% and 82.7\%, respectively, with five samples from each novel class.
\item A collaborative sensing approach is proposed to leverage the spatial diversity gain when there are multiple receivers. The classification accuracy can be boosted to $100\%$ with six receivers, when it is tested using the Widar dataset.
\end{itemize}

The rest of this paper is organised as follows. 
Section~\ref{statement} shows the CSI model in the Wi-Fi sensing system and states the problem of the existing system. 
Then we present the overview of our system design in Section~\ref{systemOverview}. 
Section~\ref{sec:method} introduces our metric-based FSL method in details and 
Section~\ref{sec:collab} presents our collaborative sensing model to enable multiple receivers to work collaboratively. The performance evaluation of FewSense is given in Section \ref{sec:exper}. 
We review innovative works of FSL and cross-domain Wi-Fi sensing works in Section \ref{relate}. 
Finally, Section \ref{conclusion} concludes the paper.

\section{Background and Motivation}\label{statement}
		
\subsection{Background}\label{sec:system_model}
\begin{figure}[!t]
\centering
\includegraphics[width =3.4in]{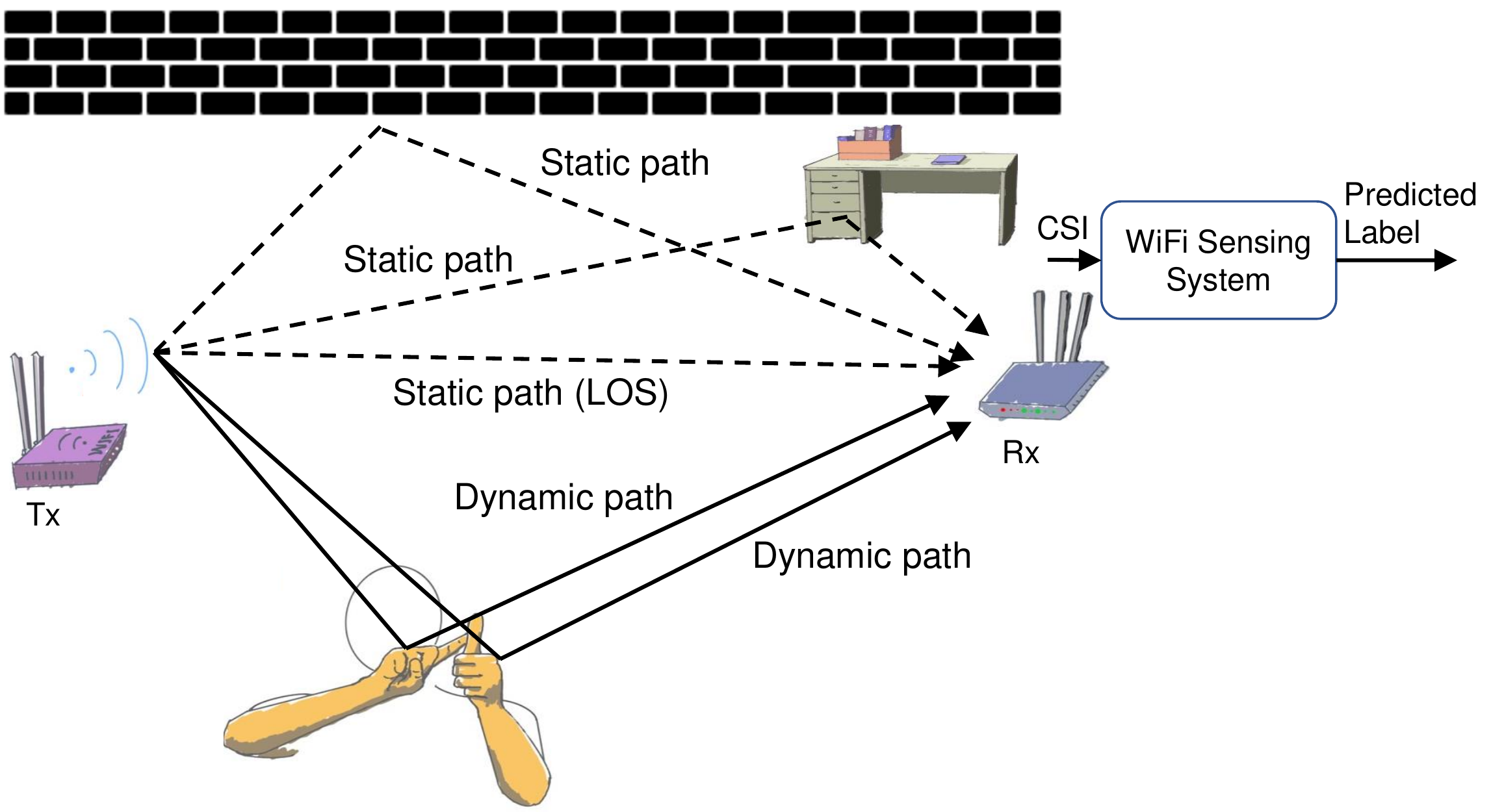}
\caption{Illustration of Wi-Fi sensing. Gesture recognition is shown as an example.}
\label{fig:system_model}
\end{figure}

As shown in Fig.~\ref{fig:system_model}, when a signal is sent from a transmitter and captured by a receiver, it experiences various propagation paths in a multipath environment. Specifically, there are static paths, which may include line-of-sight (LOS) and signals reflected by static objects such as walls and furniture, e.g., tables. There are also dynamic paths, which are reflected and/or scattered by moving objects in the environment~\cite{wang2015understanding}. Taking gesture recognition as an example, the movement of the palm will lead to dynamic paths of signal propagation.

The channel model can be mathematically given as
\begin{align}
	h(\tau,t) = \sum_{s \in {S}} h(\tau_s,t) \delta(\tau - \tau_s) + \sum_{d \in {D}} h(\tau_d,t) \delta(\tau - \tau_d),
\end{align}
where $h(\tau_s,t)$ and $h(\tau_d,t)$ are the channel attenuation of the static (grouped as $\{\mathcal{S}\}$) and dynamic paths (grouped as $\{\mathcal{D}\}$), respectively, and $\delta(\cdot)$ is the Dirac delta function.
In the context of Wi-Fi sensing,  orthogonal frequency-division multiplexing (OFDM) is the physical layer modulation used for IEEE 802.11a/g/n/ac/ax, which can obtain CSI in the frequency domain, given as
\begin{align} 
\label{eqn:csimodel}
	 H(f, t)  = \sum_{s \in {S}} h(\tau_s, t) e^{-j 2 \pi f \tau_{s}} + \sum_{d \in {D}} h(\tau_d, t) e^{-j 2 \pi f \tau_{d}}.
\end{align}
The CSI variations over time are directly affected by the objects and environments. Different gestures will have different moving patterns, which lead to unique CSI variations.

Deep learning can be adopted to reveal the relationship between the CSI variations and the corresponding gestures. By collecting a series of continuous packets when a gesture is performed, we can estimate and store records of CSI. During the training stage, the deep learning model can learn the specific patterns related to different gestures. Then in the testing stage, the receiver will obtain a collection of CSI and infer the gesture based on the pre-trained deep learning model.

\subsection{Motivation}
Deep learning has been widely adopted for Wi-Fi sensing and achieved excellent classification performance. Convolutional neural networks (CNN) \cite{signfi,ma2021location}, and long-short term memory (LSTM)~\cite{zhang2020data,Yang2019LearningGestures} can learn and reveal complex feature patterns in a supervised learning manner. However, they are still subject to novel class recognition and cross-domain sensing. 

\subsubsection{Limited Scalability/Novel Class Recognition}
In real-life applications, it is common to add new gestures to and/or remove existing ones from a gesture recognition system, which will result in changes of the number of classes (gestures).
However, as already discussed, the classic deep learning techniques such as CNN and LSTM have limited scalability because they can only handle fixed-size classes once trained. Whenever new classes, e.g., unseen hand gestures, should be added, the neural network model needs to be retrained with a massive number of training samples. 
Data collection and retraining overheads thus occur~\cite{ding2020rf,xiao2021onefi}, which are time-consuming and laborious.

\subsubsection{Cross-Domain Sensing}
In order to achieve a good classification performance, the training and test datasets should share the same data distribution. Most existing works let the same user perform the gestures in the same environment \cite{wang2016wifall,wang2015understanding}. In another word, the static CSI part in (\ref{eqn:csimodel}) remains the same but the dynamic part is uniquely caused by the gestures of the user. This is termed in-domain sensing; here the domains refer to the environment and the user.

However, cross-domain sensing is actually more common in real-life deployments. 
The gestures will be probably performed by different users in different environments, i.e., cross-domain. The domain shift impacts the performance of Wi-Fi sensing systems due to two reasons. 
\begin{itemize}
	\item When the environment changes, the static parts of (\ref{eqn:csimodel}) for training and testing are different, which will result in different CSI patterns, i.e., varied data distributions.
	\item Different users may perform the same gesture in slightly different ways, which will also lead to different CSI dynamic parts of (\ref{eqn:csimodel}).
\end{itemize}

\begin{figure}[!t]
\centering
\includegraphics[width =3.4in]{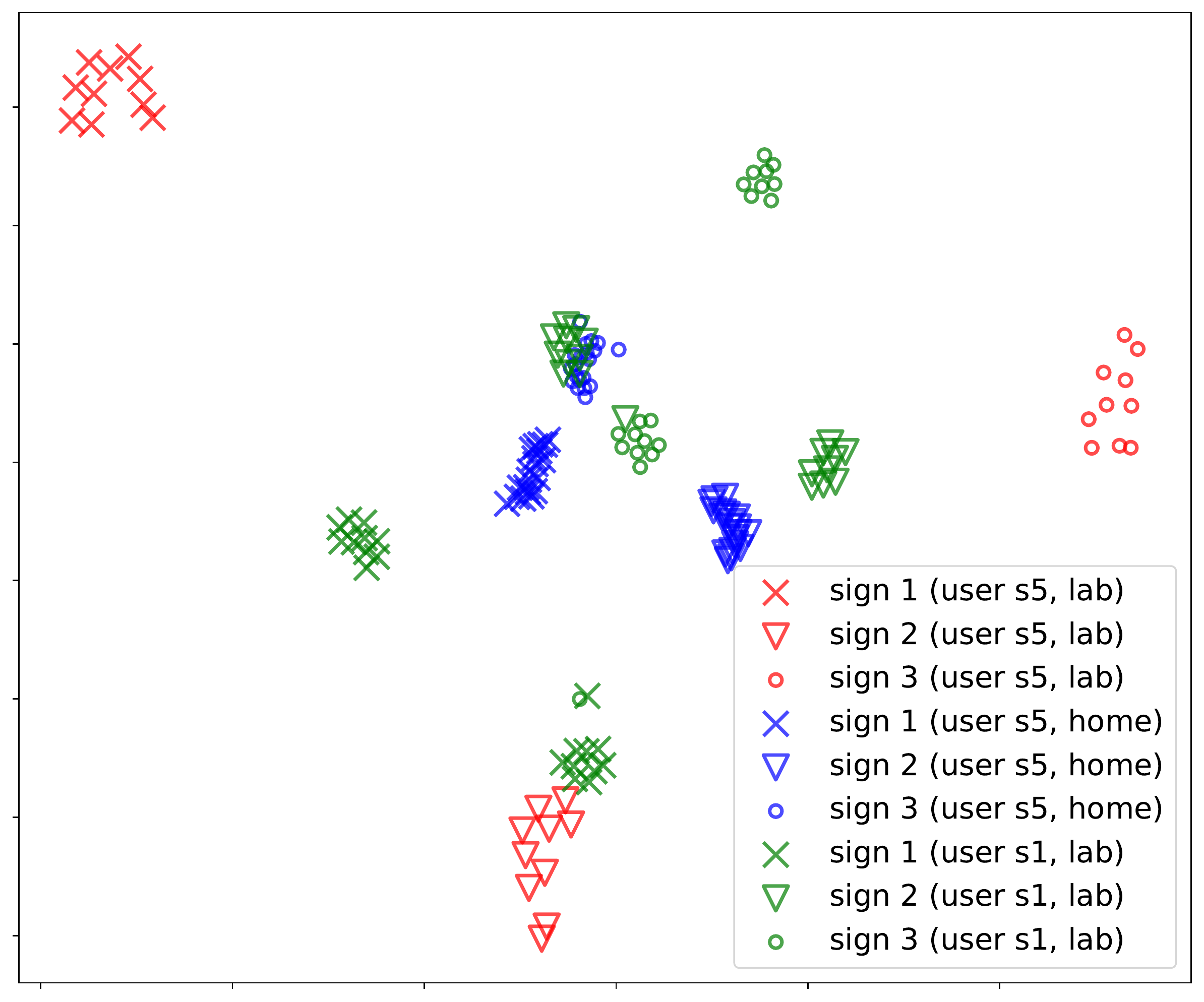}

\caption{t-SNE visualisation of the selected data from different domains: the same sign language from the same user but different environments and different sign languages from the same environment but different users.}
\label{fig:t_sne}
\end{figure}

In order to illustrate the domain shift, we use sign languages from a public dataset SignFi~\cite{signfi} that will be introduced in Section~\ref{sec:dataset}.  
Specifically, we selected three sign gestures that are performed by two users at home and in a lab environment. Their distributions are visualised using t-SNE~\cite{van2008visualizing}, which maps the CSI samples to a two-dimensional space. As shown in Fig.~\ref{fig:t_sne}, the same sign performed in different environments or by different users has totally different distributions. Therefore, directly applying the neural network model trained on one domain to a new domain would not work.

\section{System Design}\label{systemOverview}

\subsection{System Overview}
\begin{figure}[!t]
\centering
\includegraphics[width =3.4in]{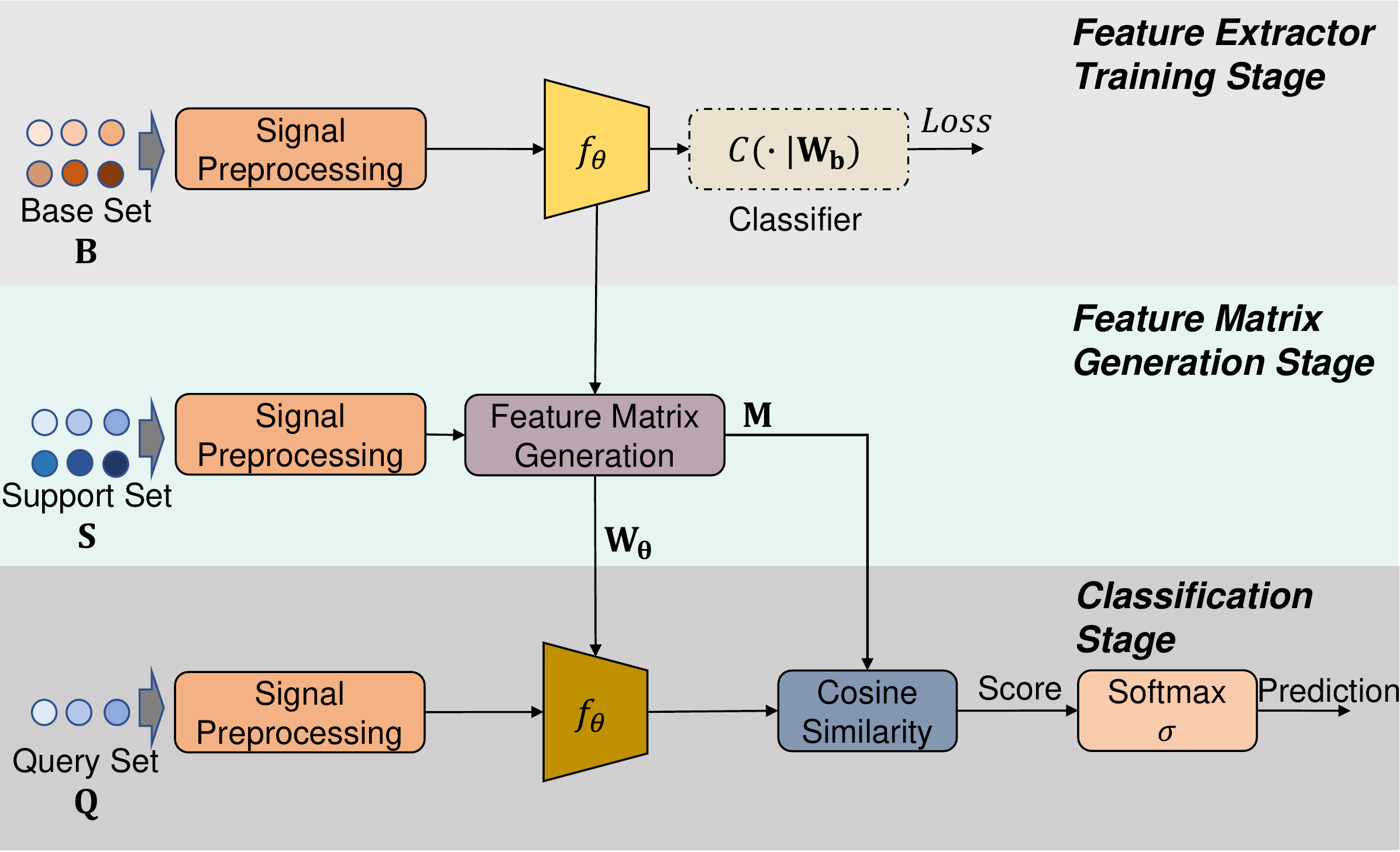}
\caption{The proposed metric-based FSL system  for Wi-Fi sensing.} 
\label{fig:system_overview}
\end{figure}

This paper employs metric-based FSL to address the above limitations. Our goal is to recognise novel classes in cross-domain scenarios, i.e., different environments and/or users, using only a few samples. Specifically, a three-stage approach is designed, namely feature extractor training, feature matrix generation, and classification, as shown in Fig.~\ref{fig:system_overview}. Note that the same signal preprocessing algorithm is used for all three stages. 

\subsection{Feature Extractor Training}
We will first sample and preprocess raw CSI records and obtain CSI tensors $\mathbf{x}$, whose construction will be elaborated in Section~\ref{sec:preprocess}.
A base set, $\mathbf{B}=\left\{\left(\mathcal{X}^{b}, \mathcal{Y}^{b}\right)\right\}$, will be constructed by obtaining CSI tensors for different classes (e.g., sign gestures),
where $\mathcal{X}^{b}=\left\{\mathbf{x}^{b}_{1}, \mathbf{x}^{b}_{2}, \cdots, \mathbf{x}^{b}_{\tilde{N}_b}\right\}$ are the CSI tensors and $\mathcal{Y}^{b}=\left\{y_{1}^b, y_{2}^b, \cdots, y_{\tilde{N}_b}^b\right\}$ are the corresponding labels.
The number of instances in the base set is denoted as $\tilde{N}_b$, and  $\tilde{N}_b = N_b \times K_b$, where $N_b$ is the number of base classes and $K_b$ is the number of samples per base class.
The domain of the base set is referred as the source domain, denoted as $\mathcal{D}_{s}$.

We use a standard supervised learning manner to train a feature extractor $f_{\theta}$ and a classifier $C\left(\cdot \mid \mathbf{W_b}\right)$, parametrised by $\mathbf{W_{\theta}}$ and $\mathbf{W_{b}}$, respectively. A large number of classes in the base set enables the feature extractor to learn unique latent features from base classes so that the distances between instances of the same class are closer while instances of different classes are further apart.
Once the training is completed, the classifier will be removed to obtain the trained feature extractor. The training only needs to be done once.

\subsection{Feature Matrix Generation}
\begin{figure}[!t]
\centering
\includegraphics[width=3.4in]{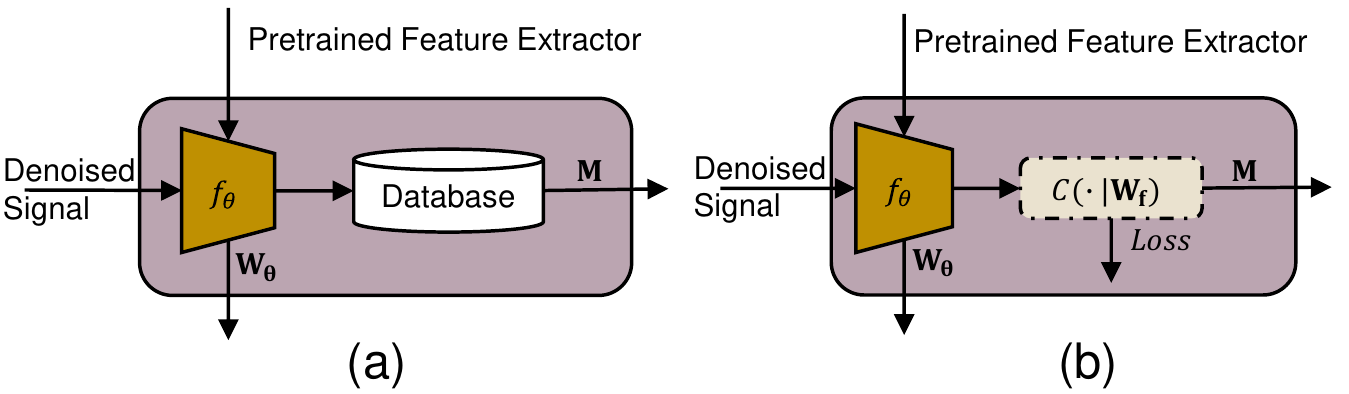}
\centering
\caption{The structure of two feature matrix generation methods.(a) Direct feature matrix generation. (b) Fine-tuned feature matrix generation.}
\label{fig:twomethods}
\end{figure}
The data involved in this stage is called the support set, which is defined as $\mathbf{S}=\left\{\left(\mathcal{X}^{s}, \mathcal{Y}^{s}\right)\right\}$, where $\mathcal{X}^{s}=\left\{\mathbf{x}^{s}_{1}, \mathbf{x}^{s}_{2}, \cdots, \mathbf{x}^{s}_{\tilde{N}_s}\right\}$ and $\mathcal{Y}^{s}=\left\{y_{1}^s, y_{2}^s, \cdots, y_{\tilde{N}_s}^s\right\}$ are the CSI tensors and labels, respectively. 
$\tilde{N}_s = N \times K$, where $N$ is the number of novel classes, and $K$ is the number of samples for each class. FSL is usually described as $N$-way $K$-shot classification.
The classes in the support set have no overlap with the classes in the base set, i.e., $\mathcal{Y}^{b} \cap \mathcal{Y}^{s}=\emptyset$. Therefore, they are called novel classes.
The domain of the support set is named as the target domain, denoted as $\mathcal{D}_t$. In practice the source domain and target domain are likely different, i.e., $\mathcal{D}_t \neq \mathcal{D}_s$, thus cross-domain sensing occurs.

We first collect $K$ labelled shots (samples) for each novel class to construct the support set, which will be denoised by the preprocessing scheme. 
With samples from the support set, the pre-trained feature extractor $f_{\theta}$ will learn the latent feature representations of the novel classes, termed as the feature matrix in this paper. We propose two methods to generate it, i.e., direct feature matrix generation and fine-tuned feature matrix generation, as shown in Fig.~\ref{fig:twomethods}.
\begin{itemize}
	\item Direct Feature Matrix Generation: The feature extractor will extract the feature embedding directly from the support set. The feature embedding is stored in the database and output as the feature matrix.
	\item  Fine-tuned Feature Matrix Generation: A new classifier will be introduced after the feature extractor. The weight matrix of the classifier and the feature extractor will be optimised using the samples in the support set. The adapted weight matrix will be the feature matrix.
\end{itemize}

\subsection{Classification }
The data involved in this stage is named as the query set 
$\mathbf{Q}=\left\{\left(\mathcal{X}^{q}, \mathcal{Y}^{q}\right)\right\}$, where $\mathcal{X}^{q}=\left\{\mathbf{x}^{q}_{1}, \mathbf{x}^{q}_{2}, \cdots, \mathbf{x}^{q}_{\tilde{N}_q}\right\}$ are the CSI tensors and $\mathcal{Y}^{q} =\left\{\mathbf{y}^{q}_{1}, \mathbf{y}^{q}_{2}, \cdots, \mathbf{y}^{q}_{\tilde{N}_q}\right\}$ represents the labels that to be inferred. 
Note that the support set and query set share the same label space and are performed in the same domain. Therefore, they have the same distribution. 

The signal preprocessing scheme is used to remove the phase noise of the query set. The feature extractor, $f_{\theta}$, is used to extract the latent features from the denoised query set. The final results are determined by computing the cosine similarity score between the latent features of the query set and the feature matrix generated from the previous stage. 

%
%

\section{Methodology}\label{sec:method}
In this section, we first introduce the signal preprocessing algorithm that is used in all three stages of the proposed FSL-based system. We then elaborate on each of these three stages. 

\subsection{Signal Preprocessing} \label{sec:preprocess}
A Wi-Fi transmitter will continuously send packets that are captured by receivers within the communication range.
The  CSI can be estimated but will be impacted by sampling frequency offset (SFO), packet detection delay (PDD) and carrier frequency offset (CFO)~\cite{xie2018precise}, which can be given as
\begin{align}
	\widehat{H}(f,t) = e^{-j\phi(t)}H(f,t),
\end{align} 
where $\phi(t)$ is the phase shift collectively caused by the above issues.

Most of the existing works only employ amplitude for sensing. The phase of CSI is more sensitive than the amplitude~\cite{wang2016rt} and will be beneficial to the system performance.
In this work, we leverage a linear transformation proposed in~\cite{wang2015phasefi} to sanitise the CSI. According to the specific implementation of the Intel 5300 Wi-Fi network interface card (NIC)~\cite{Halperin_csitool}, the process can be given as
\begin{equation}
\angle \widetilde{H}(f_i,t) = \angle \widehat{H}(f_i,t) - k m_i - b,
\end{equation}
where $i$ is the subcarrier index ranging from 1 to 30, $f_i$ is the corresponding subcarrier frequency, $m_i$ is the subcarrier index ranging from -28 to 28, and
\begin{align}
k&=\frac{\angle \widehat{H}(f_{30},t)-\angle \widehat{H}(f_{1},t)}{m_{30}-m_1},\\
b&=\frac{1}{30} \sum_{i=1}^{30} \angle \widehat{H}(f_i,t).
\end{align}


In this work, we concatenate the amplitude, $|\widehat{H}(f_i,t)|$, and the sanitised phase, $\angle \widetilde{H}(f_i,t)$. By combining the data from all the antennas, we finally construct a CSI tensor $\mathbf{x}\in \mathbb{R}^{U_s \times U_{ap} \times U_{ant}}$, where $U_s$ is the number of sampling points, $U_{ap}$ is the number of elements of amplitude plus phase, and the $U_{ant}$ is the number of antenna pairs.

\subsection{Feature Extractor Training Stage}
The main purpose of the feature extractor training stage is to train a feature extractor that can extract discriminative features from input samples. 

\begin{figure}[!t]
\centering
\includegraphics[width = 3.4in]{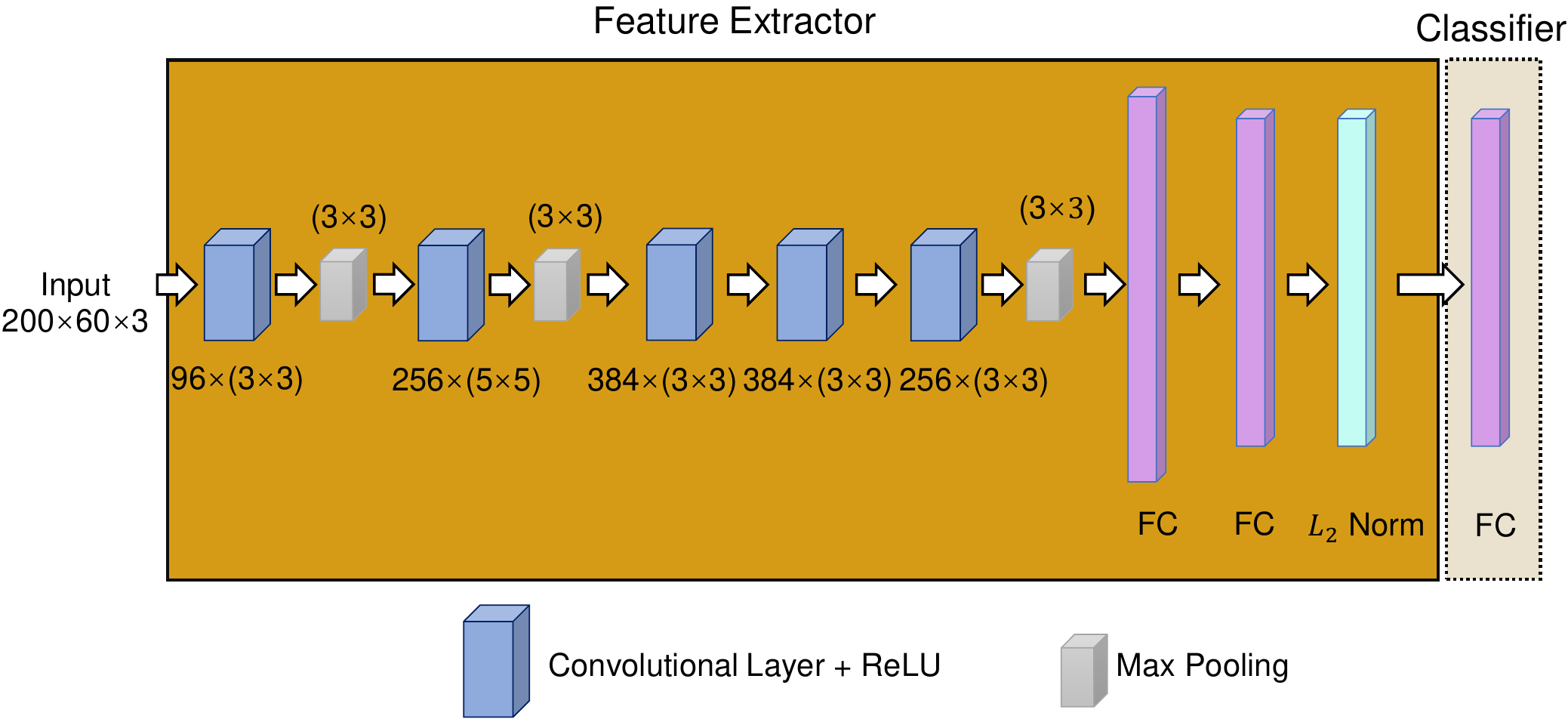}
\caption{The architecture of the feature extractor, modified AlexNet.
}
\label{fig:Alexnet}
\end{figure}

The structure of the feature extractor is shown in Fig. \ref{fig:Alexnet}, which is revised from the classic AlexNet architecture~\cite{krizhevsky2012imagenet}. An $L_2$-norm layer~\cite{ranjan2017l2} is added before the classifier, which normalises the embedded latent feature vectors as follows: 
\begin{equation}
\label{eqn:l2norm}
f_{\theta}(\mathbf{x})=\frac{f^{\prime}_{\theta}(\mathbf{x})}{\|f^{\prime}_{\theta}(\mathbf{x})\|_{2}},
\end{equation}
where  $f^{\prime}_{\theta}(\mathbf{x})$ is the output of the previous fully connected (FC) layer, and $\|\cdot\|_{2}$ denotes the $L_2$-norm operation. 
By adding an $L_2$-norm layer, the embedded features will be forced to lie on a hypersphere of a fixed radius~\cite{ranjan2017l2}. Adopting $L_2$-norm layer will improve the system's converging speed and accuracy, because it forces the embedded features from the same class closer and moves the features from different classes further apart.

A classifier, $C\left(\cdot \mid \mathbf{W}\right)$, consists of a fully connected layer with softmax activation function. In this stage, the classifier is parametrised by the weight matrix $\mathbf{W_{b}}\in \mathbb{R}^{n \times N_b}$, where $n$ is the dimension of extracted features from the feature extractor. The classifier takes the normalised feature and then computes $(\mathbf{W_b})^{\top} f_{\theta}\left(\mathbf{x}^{b}_{i}\right)$, where $(\cdot)^{\top}$ denotes transpose operation. Finally, the prediction is made via the softmax function, mathematically given as
\begin{equation}\label{eqn:cos}
\sigma(\mathbf{v})_{i} = \frac{e^{v_i}}{\sum_{j=1}^{G} e^{v_{j}}},
\end{equation}
where $\mathbf{v}$ is the output of the fully connected layer, and $G$ is the number of classes. Here, $\mathbf{v} = (\mathbf{W_b})^{\top} f_{\theta}\left(\mathbf{x}^{b}_{i}\right)$ and $G = N_b$. The output of the softmax function is a probability vector that represents confidence levels over different classes. The class with the highest probability is selected.

We train the feature extractor using the samples of the base set, $\mathbf{B}$, in a supervised learning manner. 
The feature extractor will learn to extract distinct features from the CSI samples of different classes. 
It is crucial to carry out training with a sufficient number of base classes to enable the feature extractor to generalise to novel classes. 
Once the training is completed, the classifier is removed and the feature extractor is attained.

\subsection{Feature Matrix Generation Stage}
This stage aims to generate discriminative features for novel classes in the support set. 

Firstly, one or a few samples from each novel class are obtained for the feature matrix generation. The corresponding CSI instances are added into the support set, $\mathbf{S}$. The signal preprocessing scheme will be performed on the support set data to get the input for the feature extractor.  Depending on different deployment scenarios, two methods can be used in the feature matrix generation stage.

\subsubsection{Direct Feature Matrix Generation}
The feature extractor extracts the latent feature vectors from the denoised data. The support set embedding $\mathbf{F}_{s}\in \mathbb{R}^{n \times N}$ is obtained by computing the mean of the feature vector for each class, given as
\begin{equation}
\label{supportembedding}
\mathbf{F}_{s}=\left[\begin{array}{c}
f_{\theta}{(\mathbf{x}^{s}_1)} ,
f_{\theta}{(\mathbf{x}^{s}_2}) ,
\dots ,
f_{\theta}{(\mathbf{x}^{s}_{N})}
\end{array}\right].
\end{equation}
The feature embeddings are saved in the database for future use. In this case, the output matrix $\mathbf{M}$ of this stage is the feature embedding, i.e., $\mathbf{M} = \mathbf{F}_{s}$, which will be fed into the next stage for classification. This method allows fast generating distinctive feature matrix directly from the support set with little overhead.
 
\subsubsection{Fine-Tuned Feature Matrix Generation}
In some challenging scenarios, the distributions of datasets or fundamental features may change dramatically, the feature extractor trained on the base set will fail in new sensing tasks. In order to adapt the feature extractor to new tasks,
we propose fine-tuned feature matrix generation scheme based on the fully labelled support set.

A classifier with the trainable weight matrix $\mathbf{W}_{\text{f}} \in \mathbb{R}^{n \times N}$, initialised by the support set embedding $\mathbf{F}_{s}$, is added after the pre-trained feature extractor.
The output of the softmax function is  $\mathbf{\sigma}((\mathbf{W}_{\text{f}})^{\top}  f_{\theta}{(\mathbf{x}^{s}_i)})$.

Fine-tuning is done by minimising the softmax loss based on the support set, which will update the parameters in the feature extractor $f_{\theta}$ and the weight matrix of the classifier $\mathbf{W}_\text{f}$. 

The output matrix of this method is the weight matrix of the classifier, i.e., $\mathbf{M}$ = $ \mathbf{W }_\text{f}$. The fine-tuned feature extractor will be used in the next stage. The fine-tuned feature matrix generation method enhances the adaptivity of the model in a new sensing domain or scenario.

\subsection{Classification Stage}
In this stage, the sample in the query set will be classified based on the cosine similarity score between the feature matrix and the extracted latent features of the query set.

An instance from the query set $\mathbf{Q}$ will be firstly processed by the preprocessing scheme.  The  feature extractor will extract an embedded feature vector $f_{\theta}{(\mathbf{x}^{q})} \in \mathbb{R}^{n \times 1} $. 
We compute the cosine similarity score of $f_{\theta}{(\mathbf{x}^{q})}$ pair-wisely with all the elements in $\mathbf{M}$ by
\begin{equation}\label{eqn:cos}
\mathbf{v} = \frac{\mathbf{M}^{\top}  f_{\theta}{(\mathbf{x}^{q})}}{\|\mathbf{M}\|_2\|f_{\theta}{(\mathbf{x}^{q})}\|_2}.
\end{equation}
The softmax function $\sigma$ maps the score, $\mathbf{v}$, to a probability distribution, $\mathbf{P}$, and the query set will be classified to the class with the highest probability. 

\subsection{Summary}
FSL-based sensing has a good scalability capacity. Whenever novel classes need to be added, a few samples of these classes should be collected and their feature representations can be generated with acceptable overhead.

Let us revisit Fig.~\ref{fig:t_sne} for analysing cross-domain sensing. Data of the same class from the same domain is well-clustered while data of different classes from different domains is well-separated. 
As the query set and support set are obtained from the same target domain, these samples experience the same domain shift from the base set.
By using fine-tuning, we can adapt the feature extractor and classifier weights using a few samples from the support set, i.e., a transformation from the source domain to the target domain. This approach enables the model to adapt to the target domain with minimal efforts. 

\section{Collaborative Sensing}\label{sec:collab}
In the previous sections, we have introduced a Wi-Fi sensing system using one transmitter-receiver setup, which can capture features from a particular direction. However, the position and orientation of the user cannot be predicted, hence some transmitter-receiver links may not be optimal to capture CSI variations.
The performance can be boosted by employing multiple receivers, which can enrich feature observations from different directions and work collaboratively to improve system robustness and classification accuracy. 

\begin{figure}[t]
\centering
\includegraphics[width = 3.4in]{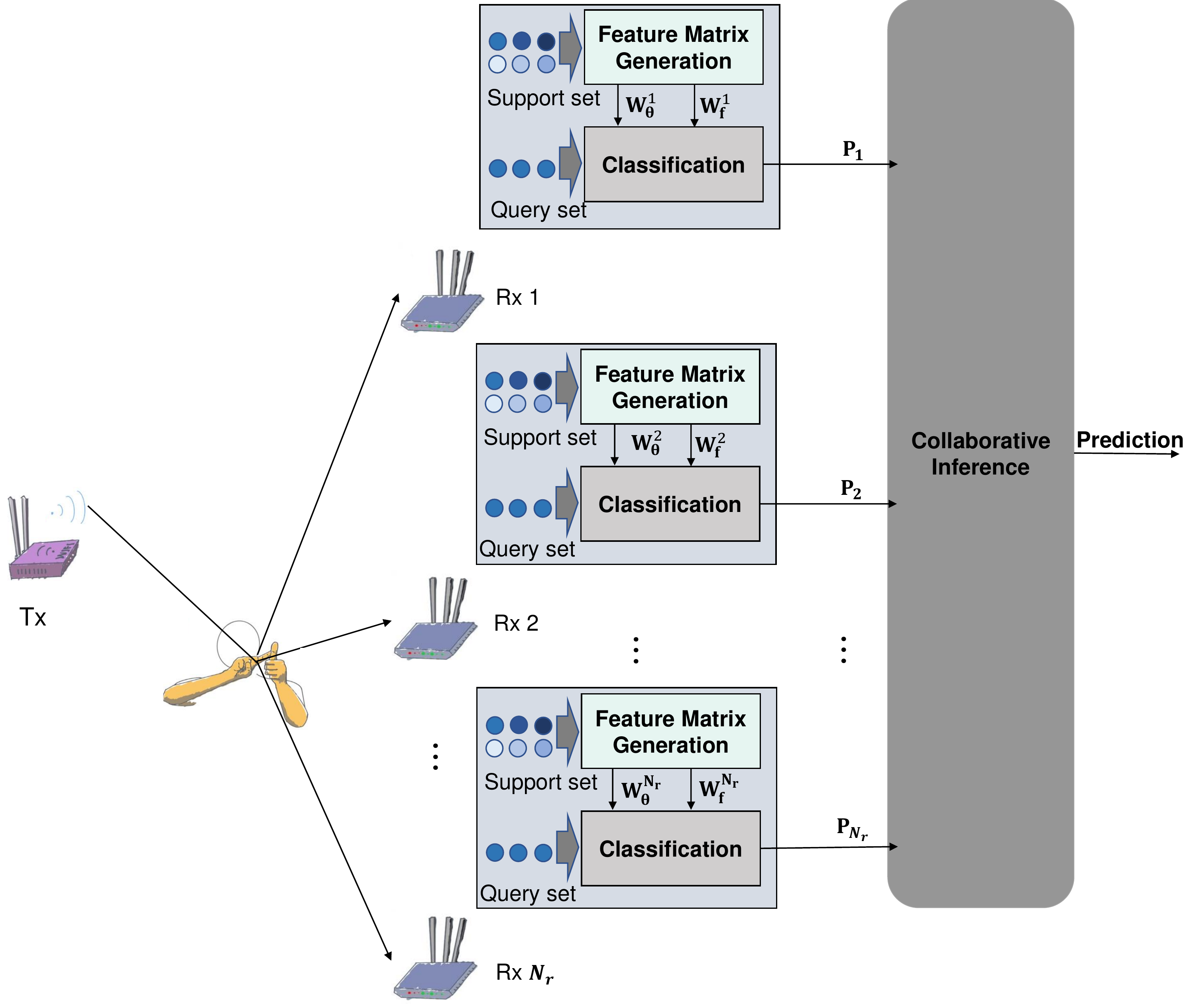}
\caption{Collaborative sensing using multiple receivers.} 
\label{fig:collabrative}
\end{figure}

As shown in Fig.~\ref{fig:collabrative}, there are $N_r$ receivers deployed for collaborative sensing. Each receiver will be equipped with the fine-tuning-FSL-based sensing approach introduced in Section~\ref{sec:method}. When a signal is sent by a transmitter, all the receivers will receive the signal simultaneously but from different directions.
Each receiver will be initially deployed with the same feature extractor.
For the $i$-th receiver, it can construct its own support set, $\mathbf{S}_{i} = {(\mathcal{X}^{s}_i, \mathcal{Y}^{s}_i)}$. 
The $i$-th receiver can then carry out fine tuning to adapt its feature extractor $f_{\theta}^{i}$ with weights $\mathbf{W}_{\theta}^{i}$ and the classifier with the weight matrix $\mathbf{W}_{\text{f}}^{i}$.

In the classification stage, each receiver will also collect its own query set, $\mathbf{Q}_i$. 
The $i$-th receiver will first carry out the independent inference and obtain a probability distribution, $\mathbf{P}_{i}$. They will then work together by calculating an averaged probability distribution of all their observations, given as
\begin{equation}\label{eqn:mojorityvote}
\mathbf{P}^{c} = \frac{1}{N_r}\sum_{i=1}^{N_{r}} {\mathbf{P}_{i}}. 
\end{equation}
The class with the highest probability is predicted as the final decision.

\section{Experimental Evaluation}\label{sec:exper}

\subsection{Setup of Experiments}
\subsubsection{Dataset} \label{sec:dataset}
We used three public datasets for  evaluation, i.e., SignFi~\cite{signfi}, Widar 3.0~\cite{widar} and Wiar~\cite{guo2019wiar}.
The CSI of all these three datasets is collected by Intel 5300 Wi-Fi NIC using the IEEE 802.11 CSI tool~\cite{Halperin_csitool}. This tool reports the CSI of 30 subcarrier groups for each packet at 20 MHz channel spacing. 
The information of the three datasets is summarised in Table~\ref{tab:dataset}. The datasets are collected by different research groups hence their volunteers and experimental experiments are different.

 For each activity, we selected $U_{s} = 200$ packets to obtain a uniform input. 
As there are 30 subcarrier groups in each packet and each subcarrier has amplitude and phase information, thus $U_{ap} = 60$.
Finally, there are three transmitter-receiver antenna pairs, i.e., $U_{ant} = 3$. 
Thus, we have the CSI tensors: $\mathbf{x}\in \mathbb{R}^{200 \times 60 \times 3}$.

\begin{table}[]
\centering
\caption{Summary of Datasets.}
\label{tab:dataset}
\begin{tabular}{|l|l|l|L{1.7cm}|L{1.3cm}|}
\hline
\multicolumn{1}{|l|}{}  & \multicolumn{1}{l|}{\textbf{Environment}} & \textbf{User ID} & \textbf{\# classes $\times$    \# reps $\times$ \# Rx} & \textbf{\# samples} \\ \hline
\multirow{6}{*}{\textbf{SignFi}} & Lab                              & User s5  & 276$\times$20$\times$1                                   & 5520        \\ \cline{2-5} 
                        & \multirow{4}{*}{Lab 2}           & User s1  & 150$\times$10$\times$1                                   & 1500        \\ \cline{3-5} 
                        &                                  & User s2  & 150$\times$10$\times$1                                   & 1500        \\ \cline{3-5} 
                        &                                  & User s3  & 150$\times$10$\times$1                                   & 1500        \\ \cline{3-5} 
                        &                                  & User s4  & 150$\times$10$\times$1                                   & 1500        \\ \cline{2-5} 
                        & Home                             & User s5  & 276$\times$10$\times$1                                   & 2760        \\ \hline
\multirow{3}{*}{\textbf{Widar}}  & \multirow{3}{*}{Classroom}       & User w1 & 6$\times$20$\times$6                                     & 720         \\ \cline{3-5} 
                        &                                  & User w2 & 6$\times$20$\times$6                                     & 720         \\ \cline{3-5} 
                        &                                  & User w3 & 6$\times$20$\times$6                                     & 720         \\ \hline
\multirow{6}{*}{\textbf{Wiar}}   & \multirow{6}{*}{Meeting room}       & User a1 & 16$\times$30$\times$1                                    & 480         \\ \cline{3-5} 
                        &                                  & User a2 & 16$\times$30$\times$1                                    & 480         \\ \cline{3-5} 
                        &                                  & User a3 & 16$\times$30$\times$1                                    & 480         \\ \cline{3-5} 
                        &                                  & User a4 & 16$\times$30$\times$1                                    & 480         \\ \cline{3-5} 
                        &                                  & User a5 & 16$\times$30$\times$1                                    & 480         \\ \cline{3-5} 
                        &                                  & User a6 & 16$\times$30$\times$1                                    & 480         \\ \hline
\end{tabular}
\end{table}

\textbf{SignFi dataset}. The SignFi dataset~\cite{signfi} contains CSI samples of 276 different sign language gestures, which involve head, arm, hand and finger gestures. The gestures are performed by five users and in three environments. Specifically, the user s5 performs all the 276 sign language gestures at home and in a lab environment. Users s1 to s4 also carry out a subset of the gestures in the same lab room, but the settings vary, e.g., with different laptop placements and desk and chair arrangements. Hence, the lab environment for users s1 to s4, denoted as Lab 2 in this paper, is deemed different from the lab environment for user s5. User s5 performs 276 sign languages each with 20 times in the lab and 10 times at home. Users s1 to s4 perform 150 sign languages 10 times in the lab 2 environment.

\textbf{Widar dataset}. The Widar dataset~\cite{widar} is a large dataset, and we only used part of it. Specifically, we selected six gestures, namely Push \& Pull, Sweep, Clap, Slide, Draw-Zigzag, and Draw-N. These gestures involve arm and hand movements. Six receivers are used to capture each gesture. We used data from three users, denoted as users w1, w2, and w3. Each user performs 6 gestures 20 times in a classroom environment.

\textbf{Wiar dataset}. The Wiar dataset~\cite{guo2019wiar} contains 16 different human motions, namely Horizontal Arm Wave, High Arm Wave, Two Hands Wave, High Throw, Draw X, Draw Tick, Toss Paper, Forward Kick, Side Kick, Bend, Hand Clap, Walk, Phone Call, Drink Water, Sit Down, and Squat. These activities involve torso, arm and hand movements.
We used the data of six users, denoted as a1 to a6, for our experiments. Each user performs each activity 30 times in a meeting room.

\subsubsection{Training Configuration}
The feature extractor was trained in a supervised learning manner. The data of user s5 lab environment of the SignFi dataset was used as the base set. We used the Adam optimiser with an initial learning rate of $1e^{-3}$ to minimise the cross-entropy loss function. The learning rate was multiplied by 0.1 whenever the validation loss stopped decreasing for 20 epochs. The training process of the feature extractor terminated when the validation loss stopped decreasing for 50 epochs. All experiments were run on a PC with i7-8700K 3.7 GHz CPU, 16 GB memory with NVIDIA GeForce GTX 2080Ti.  Tensorflow and Keras were used.

\subsubsection{Evaluation Method}
This paper evaluated the in-domain sensing (Section~\ref{sec:indomain}), cross-domain sensing (Section~\ref{sec:crossdomain}), cross-dataset evaluation (Section~\ref{sec:crossdataset}), and collaborative sensing (Section~\ref{sec:collaborative}). 
The base set was constructed by randomly selecting $N_b$ classes each with $K_b$ samples from the user s5 lab environment in the SignFi dataset. The feature extractor, denoted as $f_\theta^{N_b}$, was trained from scratch using the base set.

We randomly selected $N$ different classes (ways) each with $K$ samples (shots) from SignFi, Widar, or Wiar datasets to construct the support set $\mathbf{S}$. The remaining data of each selected class was used as the query set. 

Classification accuracy and confusion matrix were used as the metrics.
The accuracy was defined as the number of correct predictions divided by the total predictions. The confusion matrix was used to show the number of correct and incorrect predictions.

\subsection{In-Domain Sensing}\label{sec:indomain}
In-domain sensing is evaluated in this section, i.e., the base set, support set and query set share the same domain.
Intuitively, a feature extractor will have better generalisation capability when more base classes are available.
We used the data from the user s5 collected from the lab environment in the SignFi dataset as the base set. We trained feature extractors, $f_\theta^{N_b}$, with $N_b$ base classes. Specifically, we studied $N_b = $ 50, 100, 150, 200, and 250. These $N_b$ classes were randomly selected from all the available classes, i.e., 276 classes in total. 
The rest of the classes of the user s5 lab environment were used to evaluate the feature extractor's novel class recognition performance. 
Fine-tuning was not applied.
We evaluated different number of ways, i.e., $N$ = 2 to 26, and one shot, i.e., $K$=1. 

\begin{figure}[!t]
    \centering
    \includegraphics[width = 3.4in]{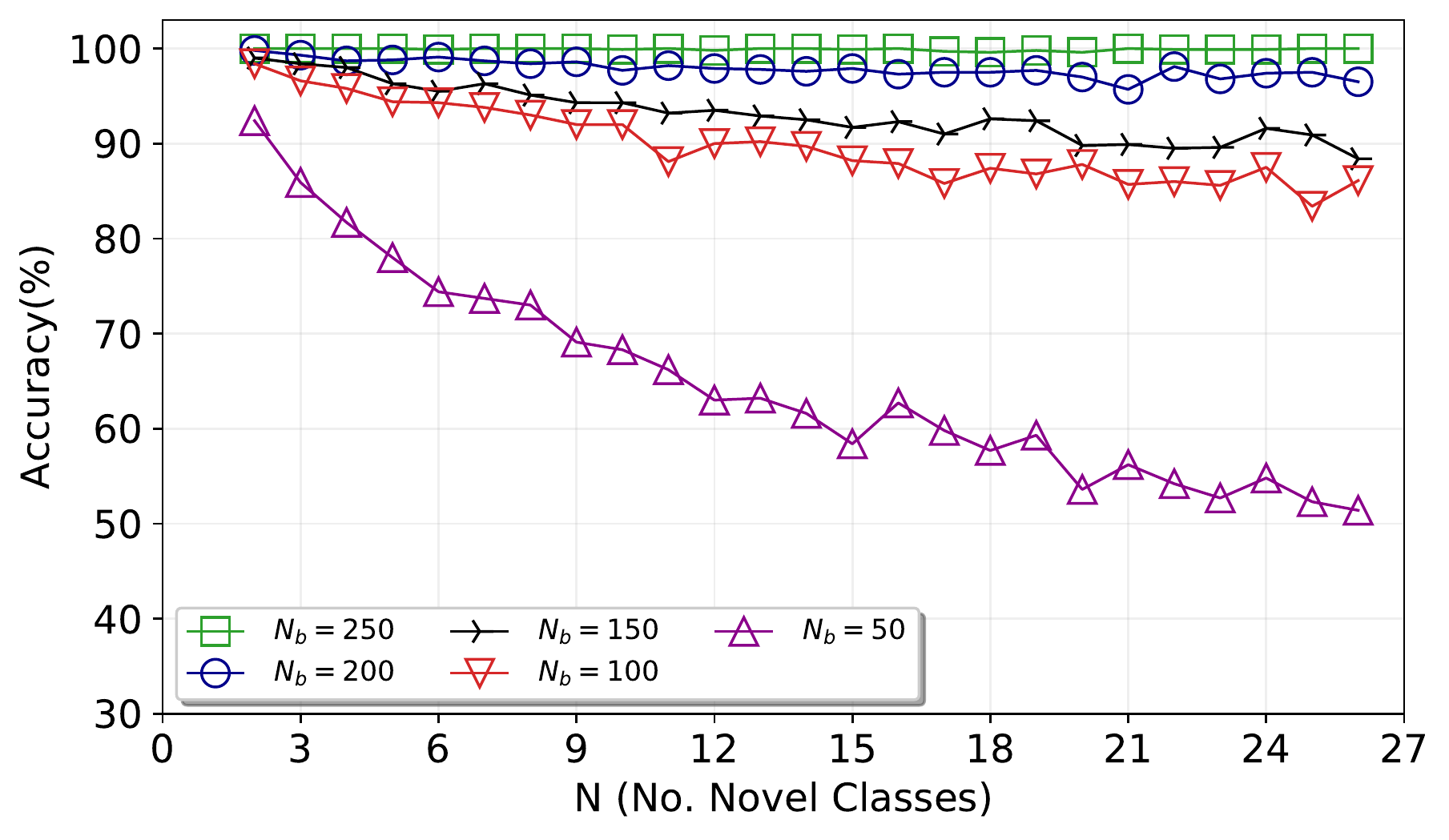}
    \caption{Performance of in-domain sensing. Feature extractors are trained with different numbers of base classes.} 
    \label{fig:comparebasedclasses}
\end{figure}

As shown in Fig. \ref{fig:comparebasedclasses}, a larger $N_b$, i.e., more base classes, leads to a higher accuracy, which is expected. When there are more novel classes in the support set, i.e., a higher $N$, the accuracy is decreasing, because it requires the feature extractor with better generalisation capability. Specifically, the average accuracies over all different numbers of novel classes are 97.9\% and 99.2\%  when $N_b = 200$ and $N_b = 250$, respectively.

In the rest parts of this paper, we used $f_\theta^{200}$ as the feature extractor. To further evaluate the performance of the feature extractor $f_\theta^{200}$, we increased the number of novel classes to 76. As shown in Table~\ref{tab:user5InLab}, the classification accuracy is $91.7\%$ when $N=$ 76, which is still quite high.

\begin{table}[]
\centering
\caption{In-Domain Sensing Performance of the Feature Extractor $f_\theta^{200}$.}
\label{tab:user5InLab}
\begin{tabular}{|l|l|l|l|l|l|l|}
\hline
\textbf{\# ways} & \textbf{30} & \textbf{40} & \textbf{50} & \textbf{60} & \textbf{70} & \textbf{76} \\ \hline
\textbf{Accuracy} & 95.4\%             & 95.8\%             & 95\%               & 93.3\%             & 92.3\%             & 91.7\%             \\ \hline
\end{tabular}
\end{table}

\subsection{Cross-Domain Sensing}\label{sec:crossdomain}
In this section, we evaluated the cross-domain sensing performance. The feature extractor $f_\theta^{200}$ is trained with the user s5 lab environment but the support and query sets are from different domains.
We have considered two scenarios:
\begin{itemize}
	\item \textbf{Scenario 1, Cross-Environment Evaluation}: The support and query sets are from the user s5 home environment, i.e., the same user but different environments. As $N_b$ = 200 classes are used for training the feature extractor, there are 76 remaining novel classes.
	\item \textbf{Scenario 2, Cross-Environment and Cross-User Evaluation}: The support and query sets are from users s1 to s4 in the lab 2 environment, i.e., different users and different environments. In the SignFi dataset, these four users performed 150 classes of sign language and 25 of them are novel classes.
\end{itemize}

\subsubsection{Impact of Numbers of Novel Classes} \label{sec:novel_class}
We evaluated $N$-way 1-shot learning and fine-tuning was not used.
Only the data of novel classes was used, hence we evaluated $N$ = 2 to 76 and $N$ = 2 to 25 for scenarios 1 and 2, respectively.  

\begin{figure}
\centering
    \subfloat[]{\includegraphics[width=3.4in]{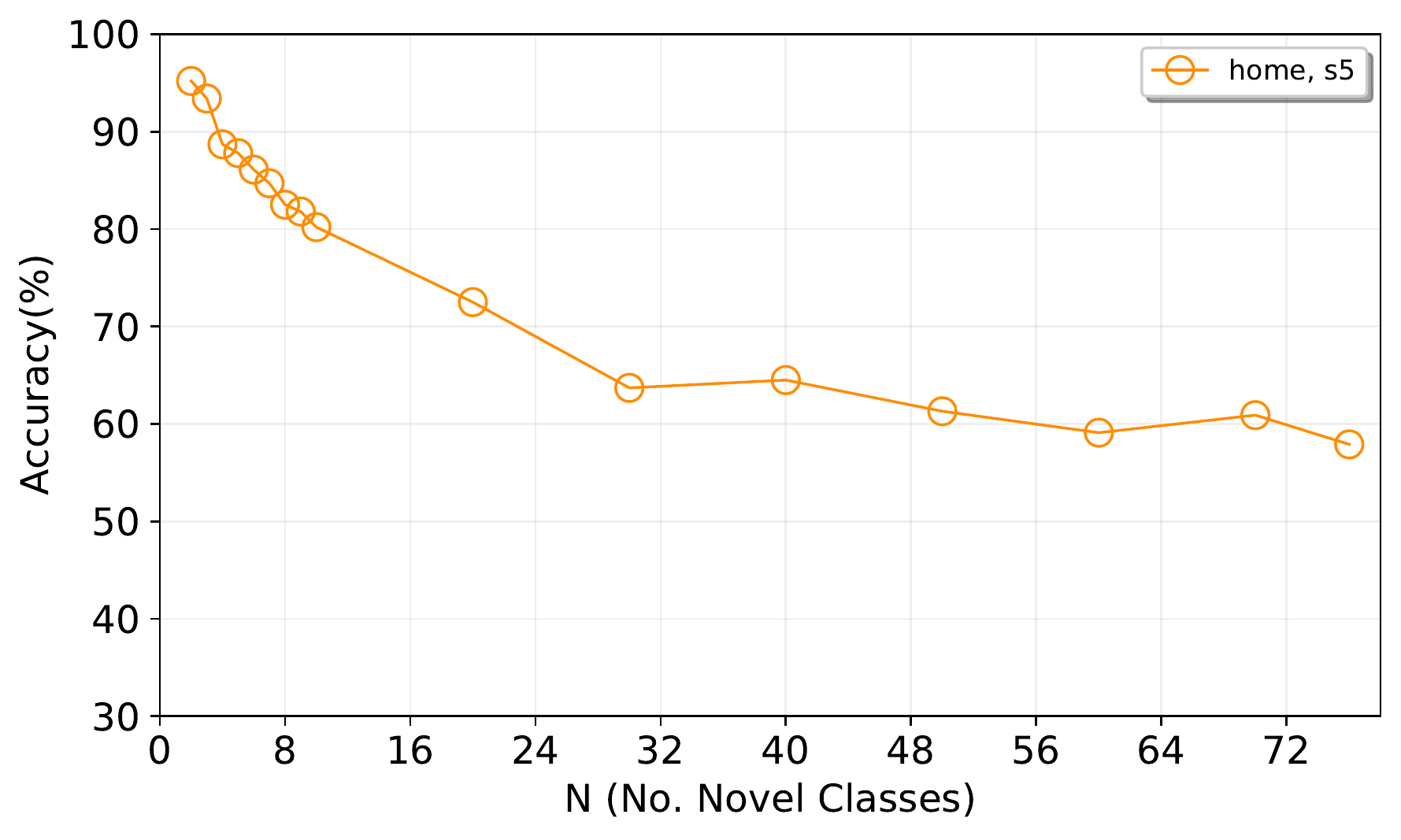}
    \label{fig:CrossDomainSameUser}}
    
	\subfloat[]{\includegraphics[width=3.4in]{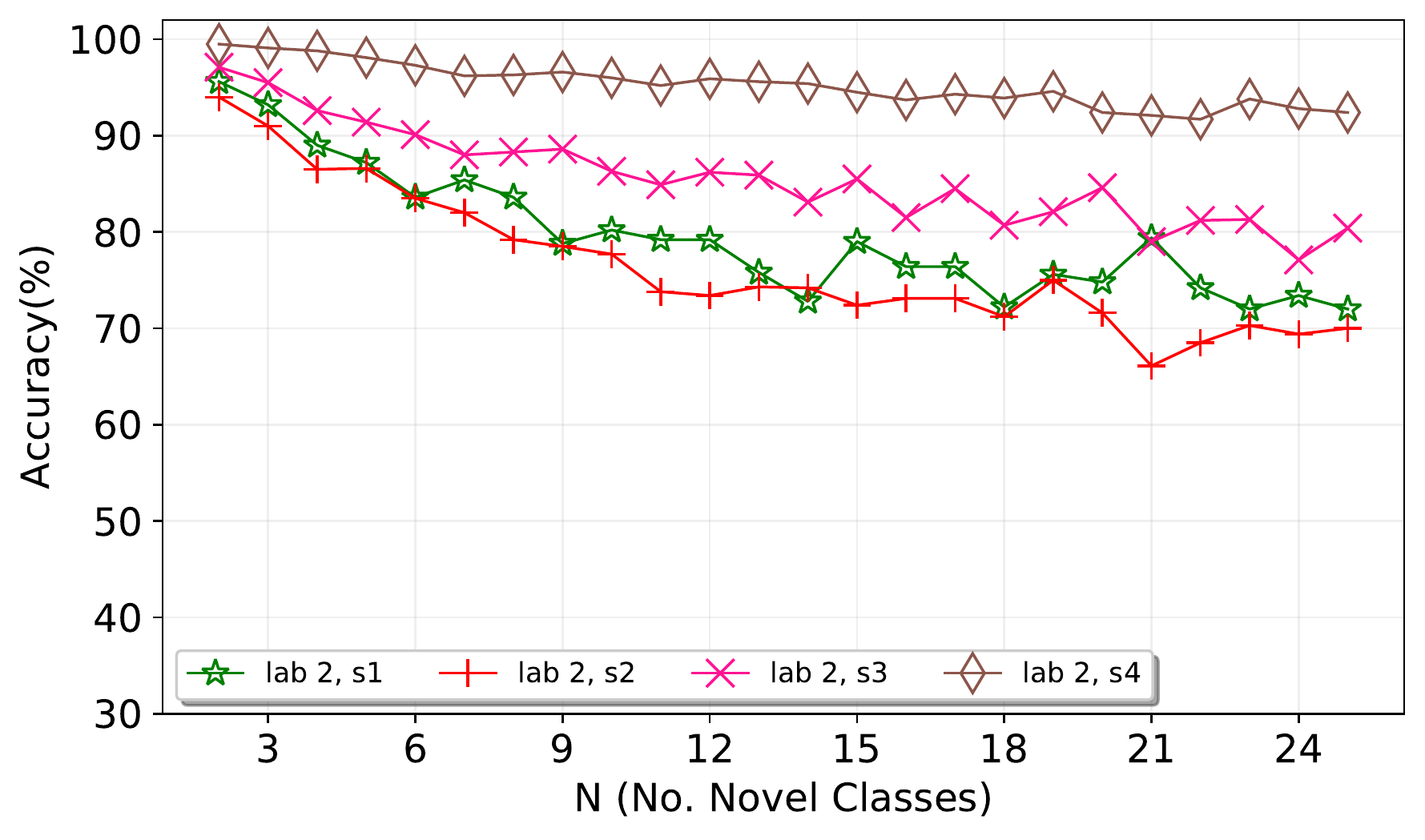}
    \label{fig:crossDiffUserandEnvironment}}
    \caption{Performance of cross-domain sensing. Impact of number of novel classes. Fine-tuning is not applied. The feature extractor $f_\theta^{200}$ is used. (a) Scenario 1, Cross-Environment Evaluation. (b) Scenario 2, Cross-Environment and Cross-User Evaluation.}
    \label{fig:NoClasses}
\end{figure}

As shown in Fig.~\ref{fig:NoClasses}, the accuracies of both scenarios decrease when the number of novel classes, $N$, is increasing, as a better generalisation capability of the feature extractor is required.
The accuracy of scenario 1 dropped to $57.9\%$ when $N$ = 76.
Regarding scenario 2, the results in Fig.~\ref{fig:crossDiffUserandEnvironment} demonstrate that the system can learn the distinctive representation of the novel classes, even though the feature extractor has not seen data from these environments and users before.
The accuracies vary for different users because each user may perform the same sign language  in a different manner, which results in different data distribution.

\subsubsection{Impact of  Numbers of Shots} \label{sec:shots}
This section studied the impacts of different numbers of shots, $K$, in the support set.  Fine-tuning was not used. We increased $K$ from 1 to 5 for both scenarios 1 and 2.

\begin{figure}[!t]
\centering
    \includegraphics[width=3.4in]{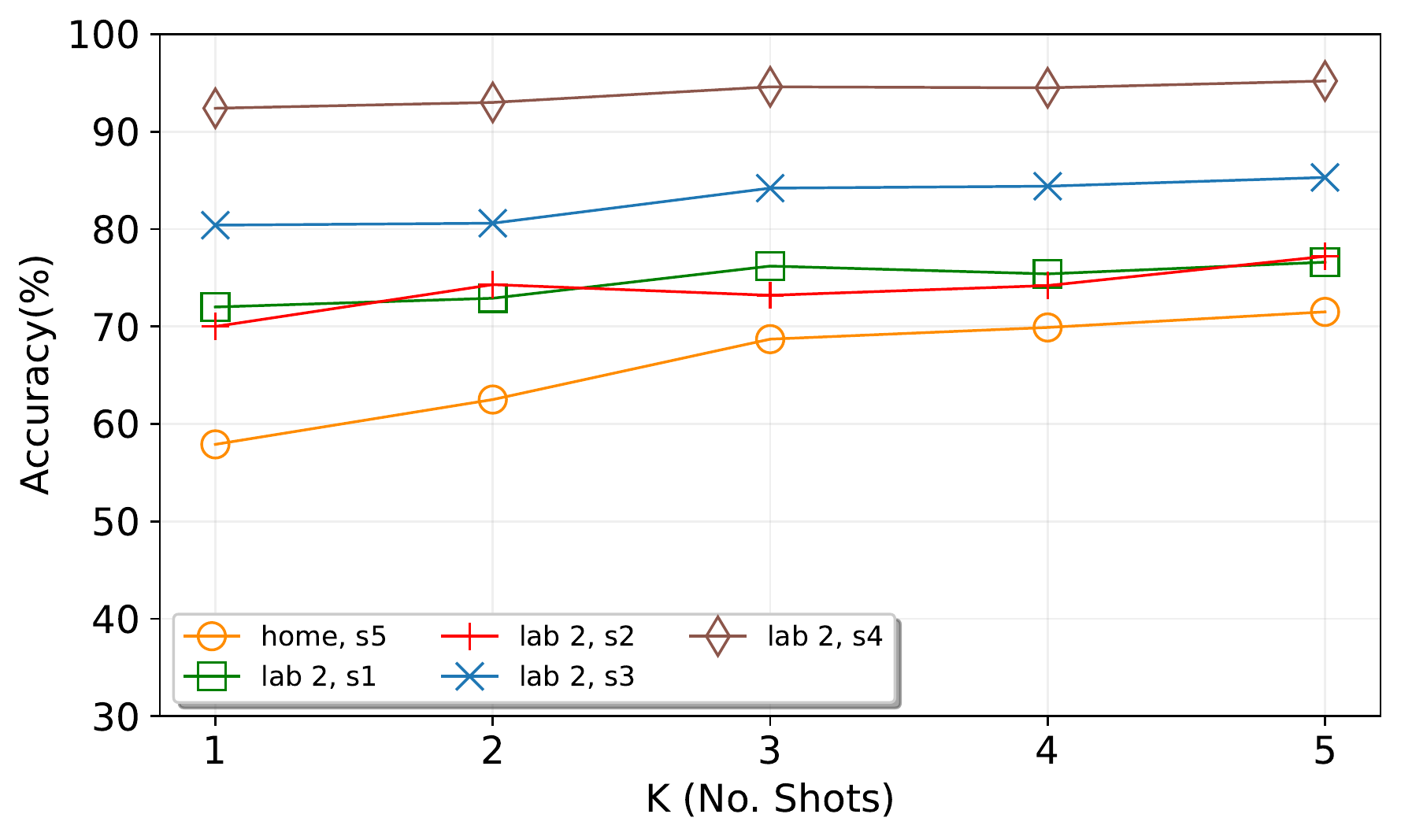}
    \caption{Performance of cross-domain sensing. Impact of number of shots in scenario 1 (user s5) and scenario 2 (user s1-s4). Fine-tuning is not applied. The feature extractor $f_\theta^{200}$ is used.} 
    \label{fig:fsl}
\end{figure}

The results are shown in Fig.~\ref{fig:fsl}. For scenario 1, we used all the remaining 76 classes, i.e., 76-way $K$-shot.
The accuracy increased from $57.9\%$ to $71.5\%$ when $K$  increased from 1 to 5, respectively. 
Regarding scenario 2, we tested 25-way $K$-shot.
Increasing the number of shots can enrich data diversity in the support set, thus increasing the generalisation capability of the feature extractor.
Again, the accuracies vary according to different users, about 20\% difference between the best and worse cases, probably due to various gesture patterns among users.
Note that the user s5 had the lowest accuracy compared to the other four users, because the number of novel classes of the user s5 was higher ($N_b$ = 76).

\subsubsection{Impact of Fine-Tuning} \label{sec:finetune}
We performed fine-tuning on the support set with only one and five samples for each novel class. We studied 76-way 1-shot and 25-way 1-shot for scenarios 1 and 2, respectively.

\begin{figure}[!t]
\centering
    \includegraphics[width=3.4in]{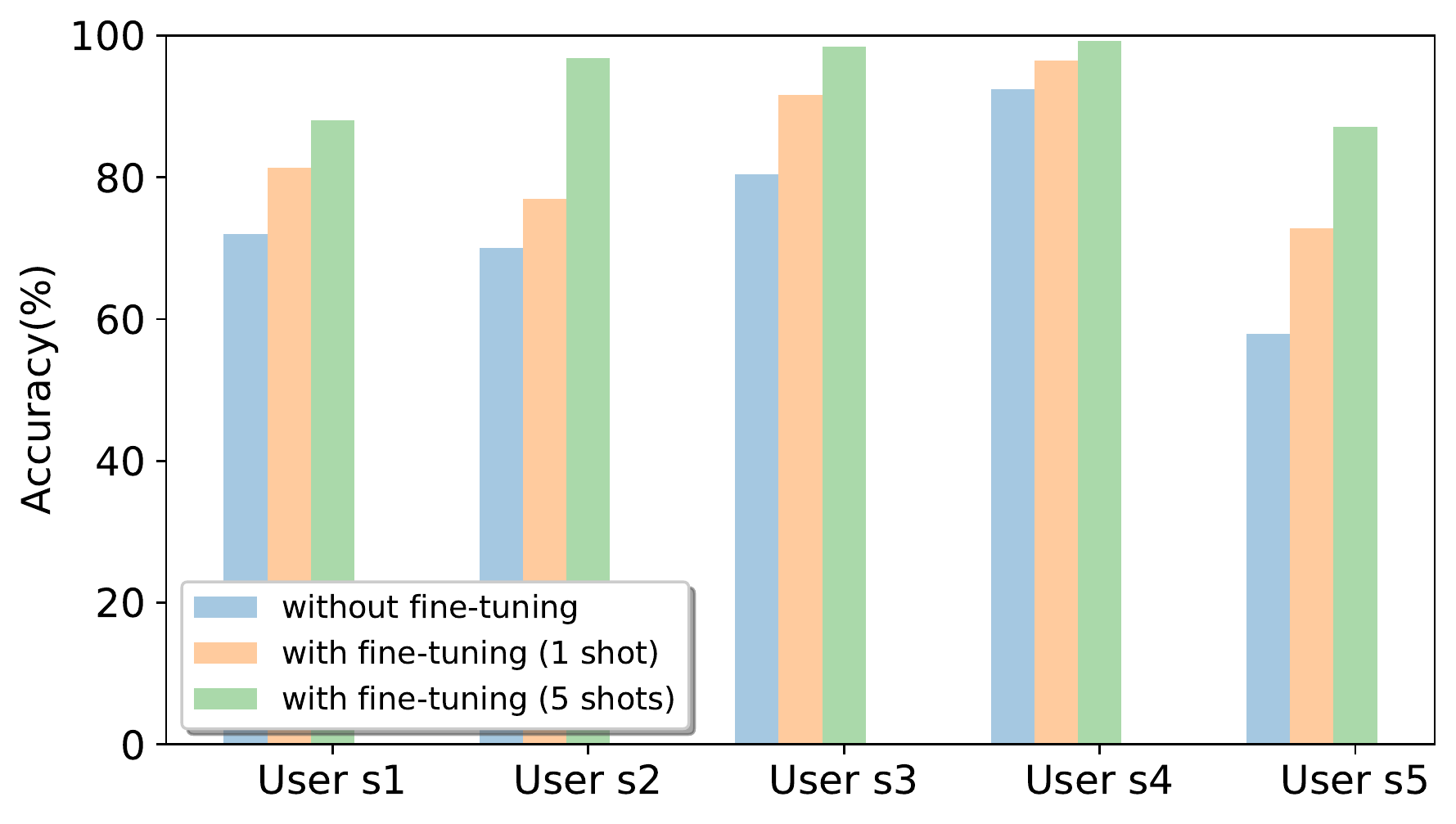}
    \caption{Performance of cross-domain sensing. Impact of fine-tuning in scenario 1 (user s5) and scenario 2 (user s1-s4). The feature extractor $f_\theta^{200}$ is used.}
    \label{fig:ft}
\end{figure}

As shown in Fig.~\ref{fig:ft}, fine-tuning has improved the accuracies for all the tests. For scenario 1 (user s5), the accuracy is improved from $57.9\%$ to $72.8\%$ and 87.1\%, when one shot and five shots are applied, respectively. For scenario 2 (users s1 to s4), the performance is improved by $8\%$ on average when one shot is applied, and improved by 16.9\% on average when five shots are applied. The evaluation results indicated that our proposed fine-tuning method could significantly improve the cross-domain's accuracy and quickly adapt to the new domain without extensive data collection, because the feature extractor can be updated even using one shot.

\subsection{Cross-Dataset Evaluation}\label{sec:crossdataset}
This section evaluated the classification performance when the base set and support \& query sets were from different datasets. The same feature extractor $f_\theta^{200}$ trained by the SignFi dataset was used.
The support and query sets were from the Widar or Wiar datasets.

This would be quite challenging. Firstly, there will be a significant domain shift between the datasets, as the users and environments are different.
Secondly, different datasets may have different sensing tasks, which result in totally different CSI variation patterns. For example, sign languages in SignFi are mainly a combination of finger, hand, and head movements,  the gestures in the Widar dataset involve hand movements, and the human motions in the Wiar dataset are large-scale human activities that involve arm, hand, limb, and leg, etc.
The tasks of SignFi and Widar are similar, but the ones between SignFi and Wiar are quite different.

\subsubsection{Evaluation on Widar Dataset}\label{sec:widar}
This section evaluated the recognition performance when the tasks of the base sets (SignFi) and the support \& query sets (Widar) were different but similar. Specifically, the gesture data in the Widar dataset was used for evaluation.

\begin{figure}[!t]
\centering
\includegraphics[width=3.4in]{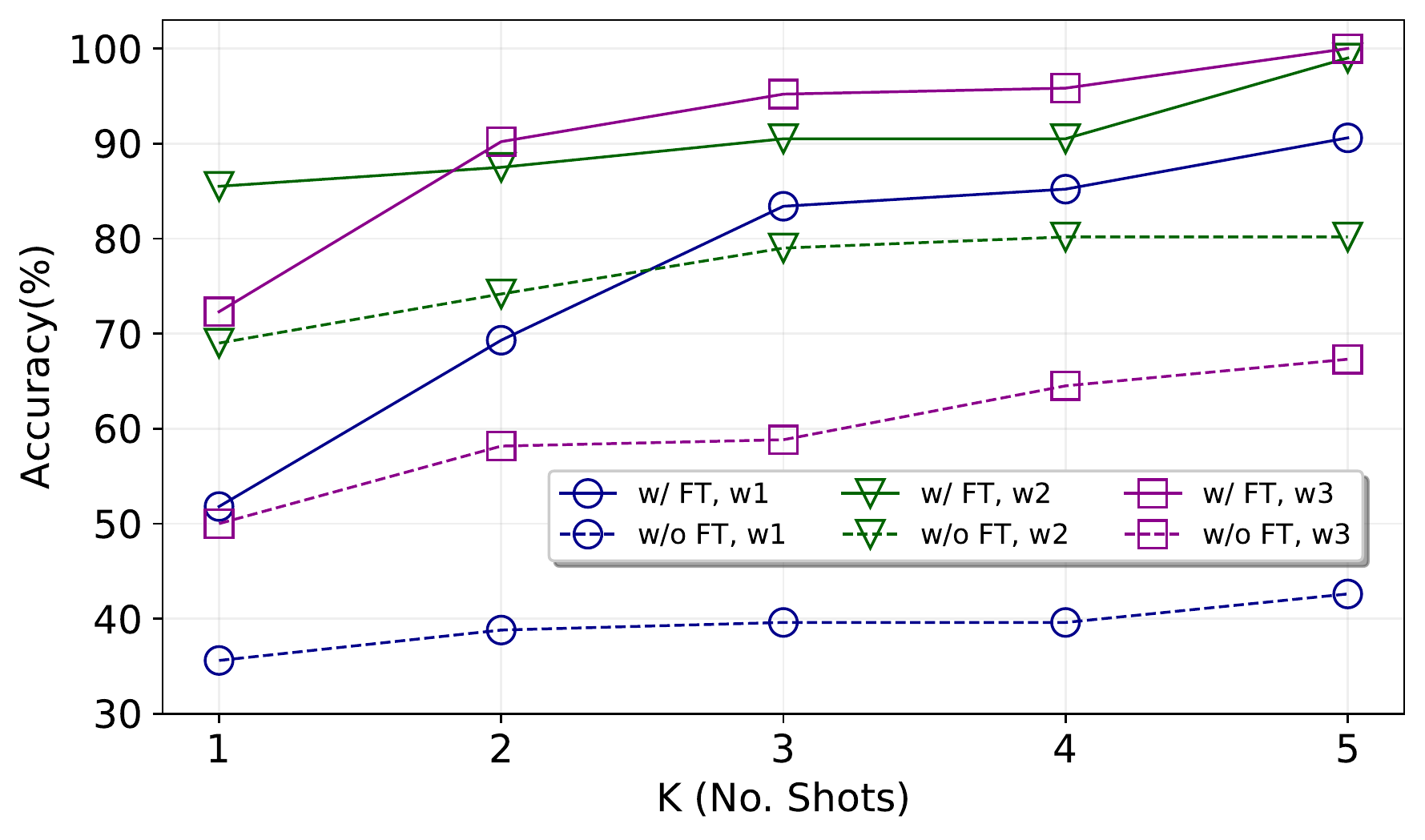}
\caption{Cross-dataset evaluation on the Widar dataset. The feature extractor $f_\theta^{200}$ is trained using SignFi dataset.}
\label{fig:widarperform}
\end{figure}

As there are multiple receivers available in Widar, we selected data of one receiver from users w1, w2, and w3. We evaluated 6-way $K$-shot, where $K$ =  1 to 5. We studied both direct feature matrix generation and fine-tuned feature matrix generation. The results are shown in Fig.~\ref{fig:widarperform}. 

\begin{figure*}[!t]
\centering
\subfloat[]{\includegraphics[width=2.2in]{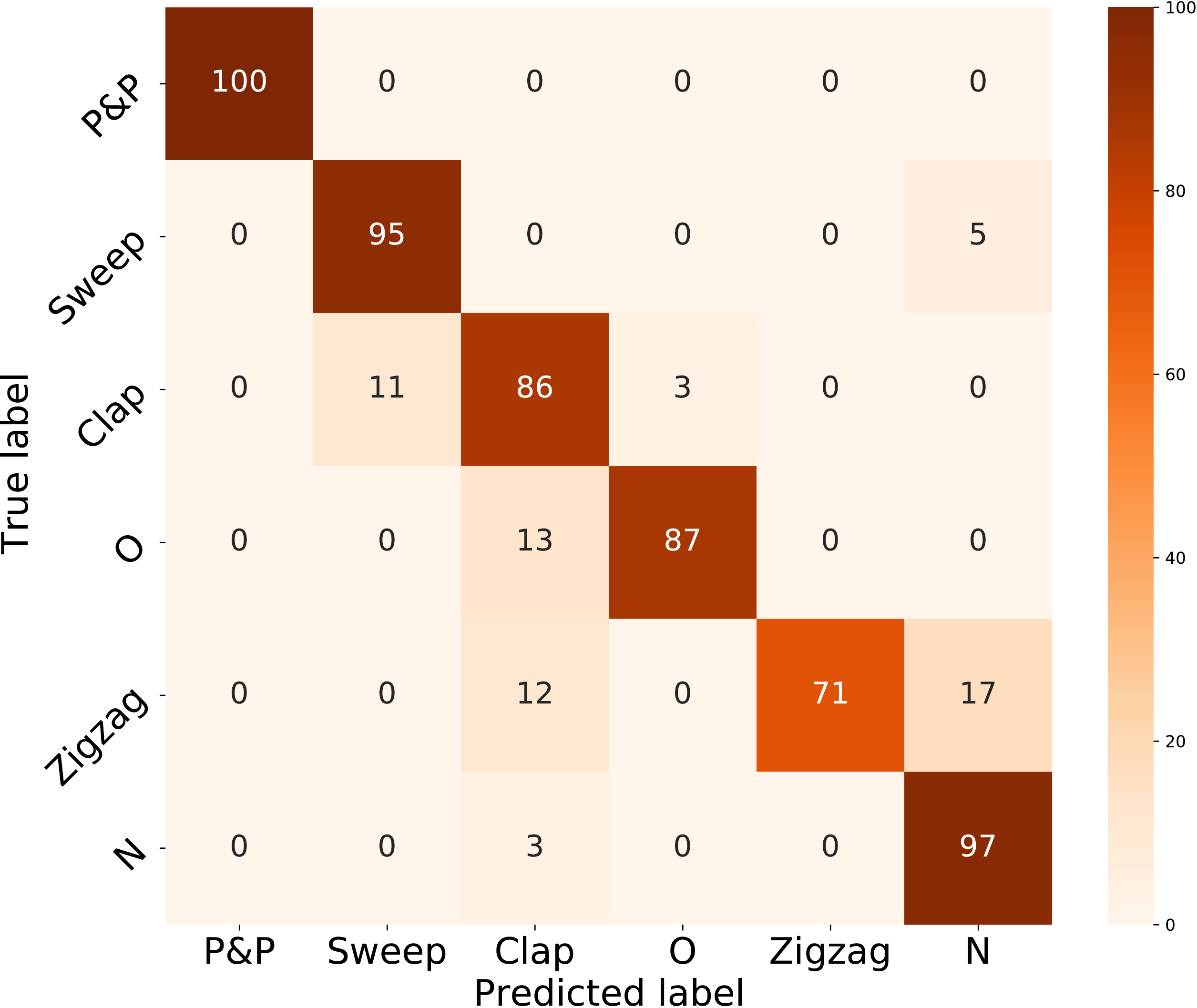}\label{fig:widar_userW1}}
\hspace{0.5cm}
\subfloat[]{\includegraphics[width=2.2in]{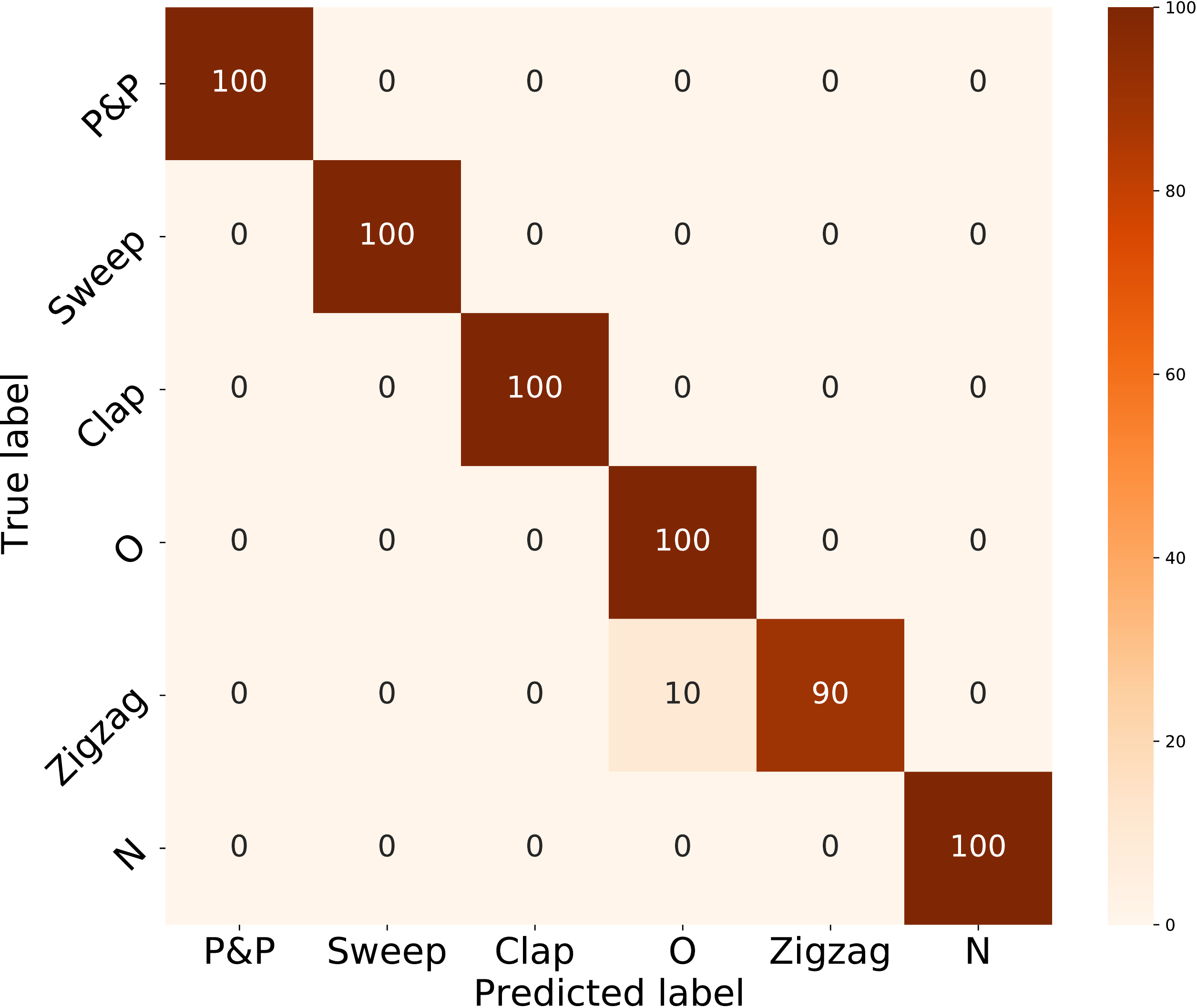}\label{fig:widar_userW2}}
\hspace{0.5cm}
\subfloat[]{\includegraphics[width=2.2in]{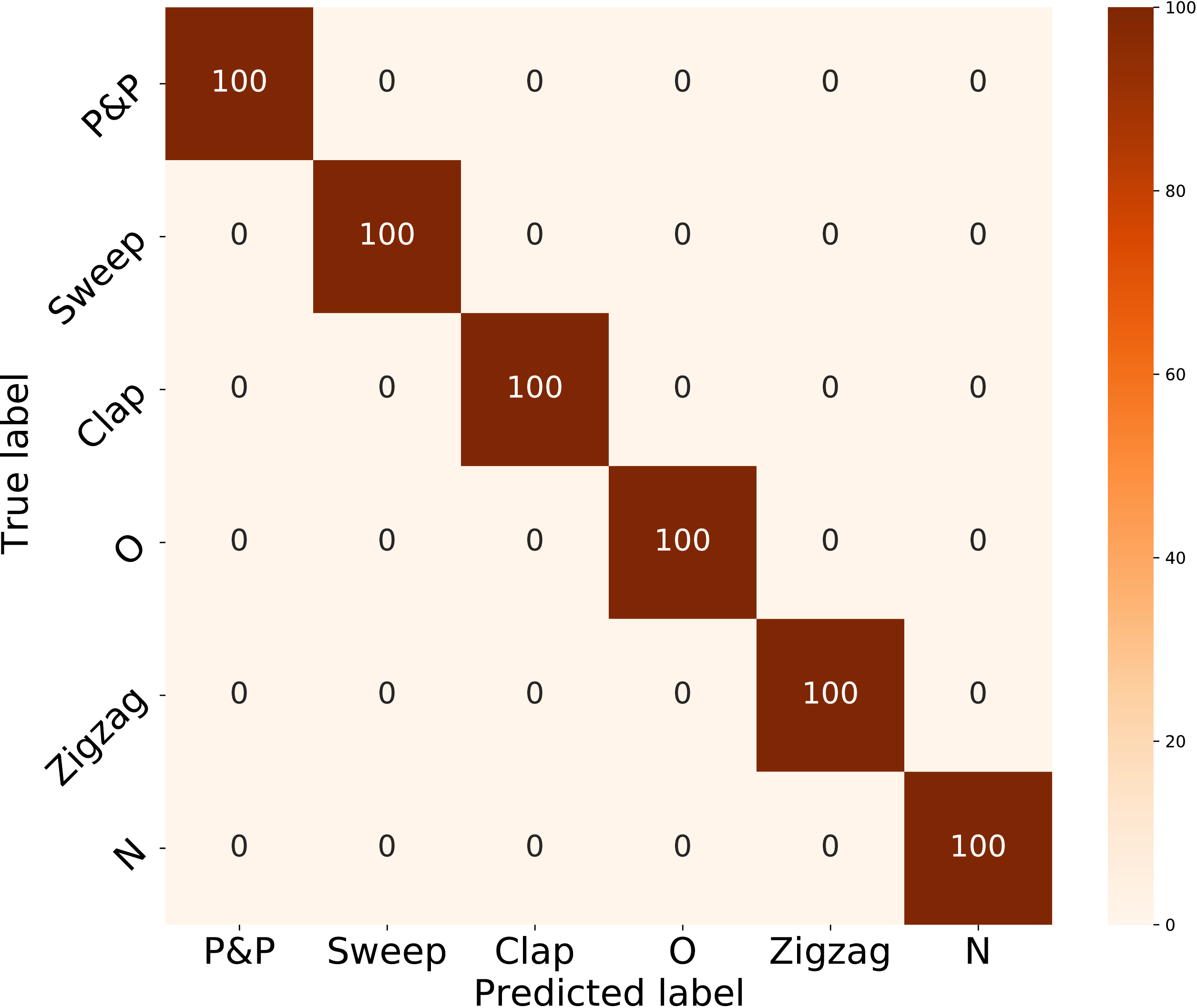}\label{fig:widar_userW3}}
\centering
\caption{Cross-dataset evaluation on the Widar dataset. The confusion matrix for 5 shots learning. (a) User w1. (b) User w2. (c) User w3.}
\label{fig:cfm}
\end{figure*}

When fine-tuning was used, the overall accuracy increased dramatically. 
The minimum accuracy increase was around $16.5\%$, when the data was from the users w1 and w2 in the one-shot learning case.
The overall accuracy increased with an increase in the number of shots. 
In the case of one-shot learning with fine-tuning, the accuracies were $51.8\%$, $86.3\%$, and $72.3\%$, for user w1, w2, and w3, respectively. The five shots learning gave the best performance, i.e., $90.6\%$, $99.0\%$ and $100\%$ for user w1, w2, and w3, respectively. The corresponding confusion matrix is shown in Fig.~\ref{fig:cfm}.
The experiment results showed that our proposed method can be scaled to similar sensing tasks, e.g., gesture recognition, with only a few samples. 

As the feature extractor was trained using the data of sign languages (SignFi), some latent features learned may not apply gesture recognition (Widar). 
The performance was thus a little limited when fine-tuning was not used.
When fine-tuning was used, even with very few samples, the recognition accuracy was significantly improved.
This is because the feature extractor has already obtained informative knowledge about how to extract latent features for sign language, and sign languages and gestures share some common features.

\subsubsection{Evaluation on Wiar Dataset}\label{sec:wiar}
This section evaluated the recognition performance when the tasks of the base sets (SignFi) and the support \& query sets (Wiar) are quite different. Specifically, the human activity data in the Wiar dataset was used for evaluation.

We selected the data from six users in the Wiar dataset. We evaluated 16-way $K$-shot, where $K$ = 1 to 5. The experiments were conducted with fine-tuning  applied.

\begin{figure}[t]
\centering
\includegraphics[width = 3.4in]{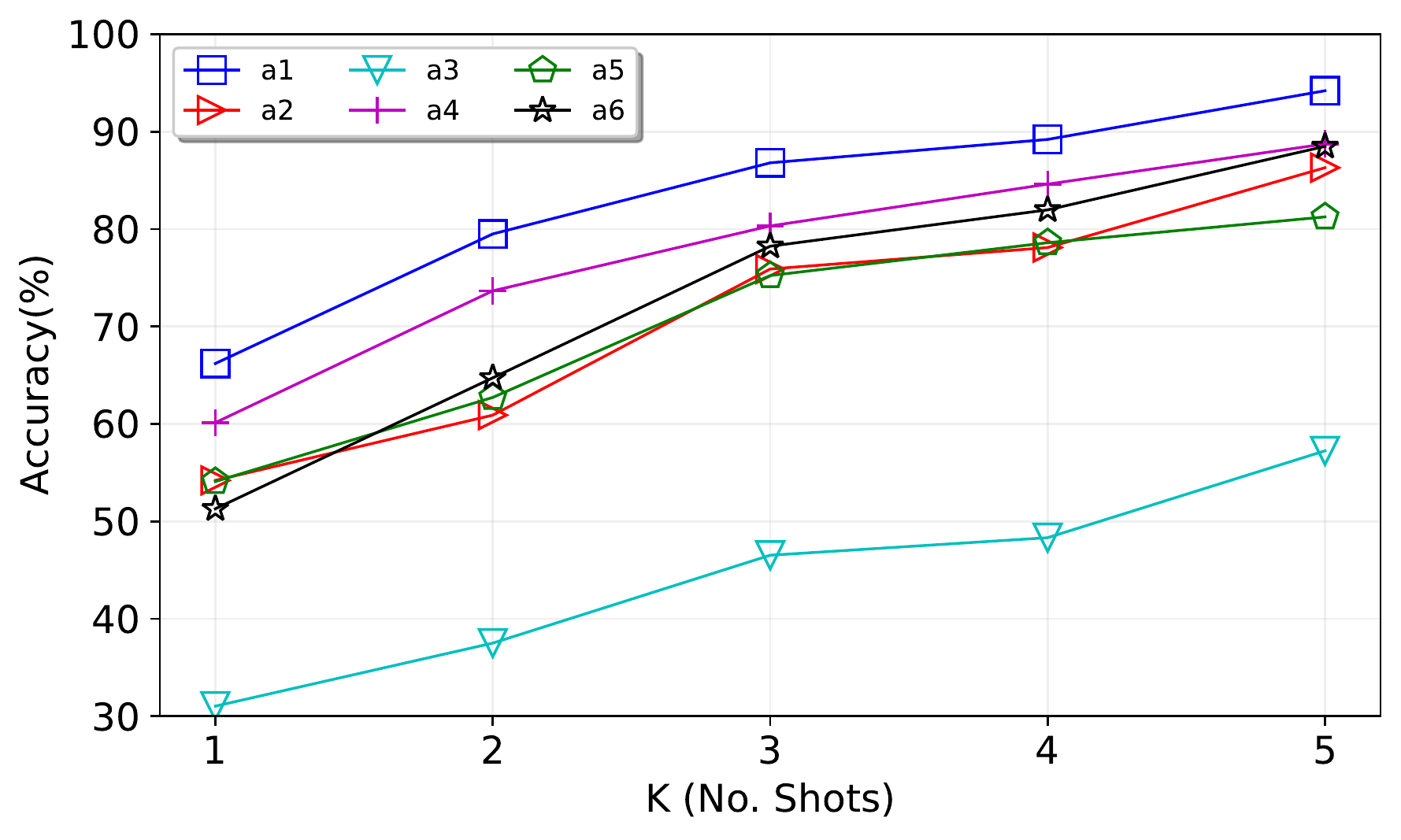}
\caption{ Cross-dataset evaluation on the Wiar dataset. The feature extractor $f_\theta^{200}$ is trained using SignFi dataset.}
\label{fig:wiar}
\end{figure}

The experiment results are shown in Fig.~\ref{fig:wiar}. The feature extractor $f^{200}_\theta$ achieved the highest accuracy for user a1, with an accuracy of $66.2\%$ and $94.2\%$ for one and five shots learning, respectively. The lowest accuracy was shown for user a3, with an accuracy of $31.0\%$ and $57.24\%$ for one and five shots learning, respectively. In the case of five-shot learning, the performance for most of the users (except for user a3) were over 80\% of accuracy.
The results demonstrate that our feature extractor could adapt to various sensing tasks without the burden of high data collection and model retraining, even when the tasks are very different to the original task. 

\subsubsection{Discussion}
The work~\cite{xiao2021onefi} also used FSL for Wi-Fi sensing and argued it is cumbersome to build a feature extractor with good generalisation capability. Hence they designed virtual gesture generation for reducing the effort of collecting data. However, as we showed in this Section, we can leverage the public dataset as the base set and train a versatile feature extractor. When the feature extractor is applied to a different sensing task, we can always fine-tune the feature extractor using very few samples from the new task. Therefore, the overhead for collecting the base set can be mitigated.

\subsection{Collaborative Sensing}\label{sec:collaborative}
In this section, we evaluated the performance of the collaborative sensing algorithm. The fine-tuning was applied in this section.
The same feature extraction, $f_\theta^{200}$, trained by the SignFi dataset was used. 
The support and query sets were from the user w1 of the Widar dataset. 
Six gestures were performed 20 times and captured by six receivers each time. Therefore, each receiver has 20 samples for each gesture. We constructed the support and query sets for each receiver. One to five samples were randomly selected and added to the support set and the remaining were used as the query set.

\begin{figure}[t]
\centering
\includegraphics[width = 3.4in]{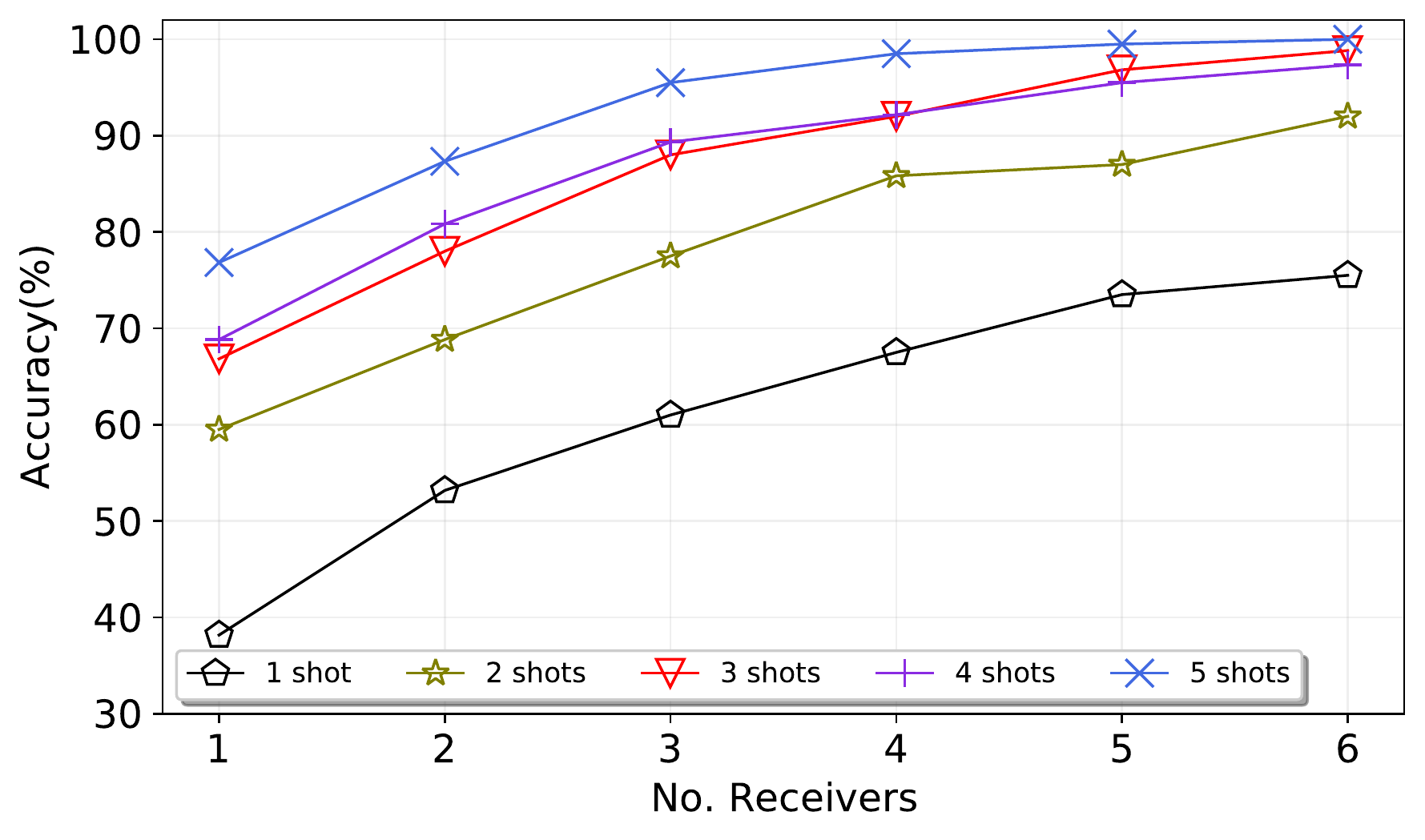}
\caption{Performance of collaborative sensing using multiple receivers. The feature extractor $f_\theta^{200}$ is trained using the SignFi dataset. Support and query sets are from Widar dataset.} 
\label{fig:multiRx}
\end{figure}

As shown in Fig.~\ref{fig:multiRx}, the accuracies increased as the number of receivers increased. For one-shot learning, the accuracy increased from $38.2\%$ to $75.5\%$ when the number of receivers increased from 1 to 6. 
Exploiting more receivers enables the feature extractor to view gestures from different perspectives, hence rich features can be obtained for improving system performance.
Similarly, the classification accuracy in collaborative sensing can also be enhanced by more shots.
The highest accuracy achieved by the 5-shot learning is $100.0\%$ when the number of receivers was six.

The work in~\cite{xiao2021onefi} also used multiple receivers. However, the accuracy in their work is already relatively high only using one receiver, hence the accuracies were saturated with more receivers. 
However, as demonstrated by our work,  more receivers will be beneficial when the accuracy using a single receiver is not saturated.

\subsection{Complexity}\label{sec:complexity}
The computational overhead involves three parts: training feature extractor, fine-tuning, and inference. The signal pre-processing algorithm used in all three stages took 0.3 seconds.

\textbf{Training feature extractor} probably is the most computationally expensive.
In order to achieve good generalisation capability, the feature extractor needs to be pre-trained with a large number of base classes. 
Specifically, we trained the feature extractor using 200 classes each with 20 samples of user s5 lab environment in the SignFi dataset, which took around 10 minutes. Fortunately, the feature extractor only needs to be trained once and can be done offline. The pre-trained feature extractor takes up 45 MB memory, which can be deployed on embedded systems.

\textbf{Fine-tuning} is applied for adapting the pre-trained feature extractor and classifier to new sensing tasks or domains. The time cost depends on the amount of data. In all the experiments, the most time consuming one is 76-way 5-shots learning, which takes 82 seconds. Thus, the overhead in this stage is acceptable. 

\textbf{Inference} is completed by the proposed collaborative sensing model when multiple receivers are available. 
The time consumption of each inference is around 0.04 seconds, which can be done almost in real-time.

\subsection{Comparison with the-State-of-the-Art}\label{sec:adv} 

\subsubsection{Comparison with Domain Adversarial Training}\label{sec:adv} 
This section compared the performance of the proposed method to a state-of-the-art domain adversarial training-based method, also know as adversarial domain adaptation, which has been employed in the Wi-Fi sensing area, e.g., \cite{jiang2018towards,kang2021context}. 

The basic idea of domain adversarial training is to train a network on data from different domains, which are termed source domains. A domain adversarial network includes a feature extractor, a classifier and a domain discriminator. The feature extractor is used to extract features. The classifier is used to identify the label of the input. The domain discriminator is used to distinguish the domain label of the input. The domain discriminator would force the feature extractor to extract domain-independent features. After training, the network is expected to perform well on target domain data.

To ensure a fair comparison, we used the same modified AlexNet architecture in this paper as the feature extractor of the adversarial network.
We trained the model with the data from two different environments, i.e., lab and lab 2, and five users of the SignFi dataset. Each environment and user pair forms a source domain. The data from the home environment, user s5, was used as the target domain data. 

The comparison results are shown in Table~\ref{tab:compareADV}.
The domain adversarial training-based approach only achieved an overall accuracy of $4\%$, which failed to recognise the sign language gestures in the target domain. This is probably because the domain adversarial network requires a large number of domains to enable the feature extractor to learn domain-independent features. For example, the work in~\cite{jiang2018towards} revealed their classification accuracy increased from about 45\% to 75\% when the number of sources domains were increased from 2 to 22.

Instead of training a feature extractor to learn the domain-independent features, our proposed method adapted to the target domain by comparing the similarity of the extracted features between target domain instances. With one sample from each class, our method achieved $57.9\%$ accuracy without fine-tuning. The accuracy can be further increased to $72.8\%$ when fine-tuning is applied. 
\begin{table}[]
\centering
\caption{Performance Comparison With Domain Adversarial Training  }
\label{tab:compareADV}
\begin{tabular}{|l|l|}
\hline
\textbf{Methods}                            & \textbf{Accuracy} \\ \hline
\textbf{FewSense without fine-tuning (1-shot)}                 & 57.9\%  \\ \hline
\textbf{FewSense with fine-tuning (1-shot)}   & 72.8\%  \\ \hline
\textbf{Domain adversarial training}               & 4.0\%      \\ \hline
\end{tabular}
\end{table}

\subsubsection{Comparison with Domain-Independent Feature-based Approach}

The work in~\cite{widar} proposed BVP, a domain-independent feature, to address the cross-domain sensing problem. 
According to~\cite{widar}, extracting BVP requires at least three receivers. The Widar dataset provides BVP   extracted from six receivers, which is used here for comparison. 

We used the same deep learning model in~\cite{widar}. In order to perform cross-domain sensing, the BVP data of users w4 to w9 collected from the hall and office environments was used for training.
The BVP data of users w1 to w3 obtained from classroom was used for evaluation.
The results are shown in Table~\ref{tab:comparewidar}. The BVP-based approach achieved an average accuracy of 82.0\%, 90.5\% and 92.4\% for the user w1, w2 and w3, respectively. 
\begin{table}[]
\centering
\caption{Cross-Domain Performance Comparison With Widar}
\label{tab:comparewidar}
\begin{tabular}{|l|l|l|l|}
\hline
\textbf{Methods}                & \textbf{User w1} & \textbf{User w2} & \textbf{User w3} \\ \hline
\textbf{Widar}                  & 82.0\%  & 90.5\%  & 92.4\%  \\ \hline
\textbf{FewSense (1-shot 6-Rx)} & 76.0\%  & 94.3\%  & 90.2\%  \\ \hline
\textbf{FewSense (2-shot 6-Rx)} & 92.0\%  & 98.3\%  & 100\%   \\ \hline
\textbf{FewSense (5-shot 2-Rx)} & 87.3\%  & 91.0\%  & 95.8\%  \\ \hline
\end{tabular}
\end{table}

In comparison, we also evaluated the performance of the FewSense method. The CSI data from the same three users collected in the classroom environment via six receivers is used as the support and query set.
As can be observed from Table~\ref{tab:comparewidar}, one-shot learning can achieve comparable performance with the BVP-based method. When two shots are used, our method outperforms the BVP.

However, the applications of BVP feature may be limited in real life since it requires at least three receivers to resolve the ambiguity problem. 
In comparison, our method provides an alternative solution when the number of receivers is limited. For example, by using five shots in the two receivers setting, our collaborative sensing model can still achieve comparable performance with the BVP features at the six receivers setting.

In terms of complexity, the BVP feature extraction is computationally expensive. For example, it took around 205 seconds to compute the BVP feature for one gesture from six receivers. The high computational complexity limits its practical use. 
On the other hand, our model is lightweight, as analysed in Section~\ref{sec:complexity}.

\section{Related work} \label{relate}
\subsection{Few-Shot Learning}

FSL has been successfully implemented in many vision-based tasks~\cite{wang2020generalizing} and the similarity-based methods are widely investigated.
Matching networks \cite{vinyals2016matching} is proposed to use a cosine similarity function to solve the FSL problem. 
The authors in \cite{chen2019closer} train a neural network with a large number of base classes and then replace the last fine-tuning layer with a cosine similarity function. 
Researchers in \cite{snell2017prototypical} propose a prototypical network (PN) for FSL, which employs Euclidean distance as the distance metric. 
The works above aim to learn transferable features and a fixed similarity function, the generalisation capability of them is limited. The work in~\cite{sung2018learning} uses a relation network with a trainable similarity metric to address the above limitation.

\subsection{Domain Robustness of Wi-Fi Sensing}
The authors in \cite{jiang2018towards,kang2021context} use an domain adversarial training-based domain adaptation method to extract domain-independent features from multiple domains.
For example, the work in~\cite{kang2021context} employs a gesture classifier to classify gestures, and a domain discriminator to recognise the domain label of the input samples. The feature extractor is trained to cheat the domain discriminator such that the domain discriminator cannot distinguish the domain labels of the input. The feature extractor is expected to map the input from different domains to the same feature space. However, this method requires a massive number of training samples from different domains to obtain satisfactory performance. 

The transfer learning focuses on leveraging the knowledge of a pre-trained model and applying the model to a different but similar task. In order to reduce the model retraining effort in the target domain, the authors in \cite{CrossSense} propose a transfer learning-based method called CrossSense, which employs an ANN-based model to translate the CSI features of gestures in the source domain to the target domain. 
Although transfer learning can reduce the data collection effort in a new domain, it still requires many data samples from the target domain to achieve a satisfying performance. Moreover, transfer learning only works when the source and target domains are similar enough. Otherwise, the problem of negative transfer may limit the applications of this method~\cite{weiss2016survey}.

The domain-independent feature-based method is another type of solution. Widar~\cite{widar} uses multiple receivers to extract BVP. However, at least three receivers are required to resolve the velocity direction ambiguity problem, and the feature extraction overhead cannot be ignored. Those factors may limit the application in real deployments. 

The work in~\cite{li2020wihf}  extracts the motion patterns as the input features which are experimentally demonstrated domain-independent. However, some similar gestures could have similar movement patterns. Therefore, the classes of the gestures need to be designed to avoid similarity.

\section{Conclusions}\label{conclusion}
This paper proposed FSL-based WiFi sensing, i.e., FewSense, to address the scalability and domain-dependent challenges. FewSense utilised a revised AlexNet architecture as the feature extractor to gain generalisation capability. 
The features of query and support sets were compared and the inference was made based on their cosine similarity score.
Collaborative sensing was designed to fuse observations from multiple receivers for boosting classification accuracy.
Three public Wi-Fi sensing datasets, including SignFi (sign language), Widar (gesture recognition) and Wiar (human activity recognition), were leveraged for evaluation. We carried out extensive evaluation using a feature extractor trained by the SignFi dataset.
The experimental results indicated that FewSense could be adapted to new sensing tasks or domains with a low data collection and computational cost. 
The FewSense with one-shot learning can recognise novel sign language gestures in SignFi with an average accuracy of 99.2\% and 84.2\% in in-domain and cross-domain scenarios, respectively. When applying FewSense to new sensing tasks, FewSense recognised novel gestures on the Widar dataset with an average accuracy of 69.9\% and 96.5\% for one-shot and five-shot learning, respectively.
It also achieved an average accuracy of 52.8\% and 82.7\% for one-shot and five-shot learning, respectively, for classifying novel human activities on the Wiar dataset.
Finally, our collaborative sensing approach can boost the classification accuracy by 30\% on average when there were six receivers. In summary, FewSense demonstrated that cross-dataset sensing is applicable. The generalisability of the feature extractor can be achieved using a publicly available dataset, alleviating the data collection overhead.

\bibliographystyle{IEEEtran}
\bibliography{IEEEabrv,main}

\end{document}